\documentclass[apj,revtex4]{emulateapj}                           
\usepackage{graphics,graphicx,amsmath}
\usepackage{rotating}
\usepackage{float}

\newcommand{\um}{${\rm \mu m}$~}
\newcommand{\mm}{${\rm \mu m}$}

\long\def\symbolfootnote[#1]#2{\begingroup%
\def\thefootnote{\fnsymbol{footnote}}\footnote[#1]{#2}\endgroup}

\slugcomment{AJ in press.} 

\begin{document}

\title{Extreme Debris Disk Variability -- Exploring the Diverse Outcomes of Large Asteroid Impacts During
the Era of Terrestrial Planet Formation}

\author{Kate Y. L. Su\altaffilmark{1},
  Alan P. Jackson\altaffilmark{2,3},
  Andr\'as G\'asp\'ar\altaffilmark{1},
  George H. Rieke\altaffilmark{1},
  Ruobing Dong\altaffilmark{4,1},
  Johan Olofsson\altaffilmark{5,6},
  G.~M.~Kennedy\altaffilmark{7},
  Zo\"e M. Leinhardt\altaffilmark{8},
  Renu Malhotra\altaffilmark{9},
  Michael Hammer\altaffilmark{1},
  Huan Y.~A.~Meng\altaffilmark{1},
  W. Rujopakarn\altaffilmark{10,11,12},
  Joseph E. Rodriguez\altaffilmark{13},
  Joshua Pepper\altaffilmark{14},
  D. E. Reichart\altaffilmark{15},
  David James\altaffilmark{16},
  Keivan G. Stassun\altaffilmark{17,18}
  }

\altaffiltext{1}{Steward Observatory, University of Arizona, 933 North Cherry Ave., Tucson, AZ 85721}
\altaffiltext{2}{Centre for Planetary Sciences, University of Toronto at Scarborough, 1265 Military Trail, Toronto, ON M1C 1A4, Canada}
\altaffiltext{3}{School of Earth and Space Exploration, Arizona State University, 781 E. Terrace Mall, Tempe, AZ 85287, USA}
\altaffiltext{4}{Department of Physics \& Astronomy, University of Victoria, Victoria, BC, V8P 1A1, Canada}
\altaffiltext{5}{Instituto de F\'isica y Astronom\'ia, Facultad de Ciencias, Universidad de Valpara\'iso, Av. Gran Breta\~na 1111, Playa Ancha, Valpara\'iso, Chile}
\altaffiltext{6}{N\'ucleo Milenio Formaci\'on Planetaria - NPF, Universidad de Valpara\'iso, Av. Gran Breta\~na 1111, Valpara\'iso, Chile}
\altaffiltext{7}{Department of Physics, University of Warwick, Gibbet Hill Road, Coventry, CV4 7AL, UK}
\altaffiltext{8}{School of Physics, University of Bristol, HH Wills Physics Laboratory, Tyndall Avenue, Bristol BS8 1TL, UK}
\altaffiltext{9}{Lunar and Planetary Laboratory, The University of Arizona, 1629 E University Boulevard, Tucson, AZ 85721, USA}
\altaffiltext{10}{Kavli Institute for the Physics and Mathematics of the Universe, The University of Tokyo Institutes for Advanced Study, The University of Tokyo, Kashiwa, Chiba 277-8583, Japan}
\altaffiltext{11}{Department of Physics, Faculty of Science, Chulalongkorn University, 254 Phayathai Road, Pathumwan, Bangkok 10330, Thailand}
\altaffiltext{12}{National Astronomical Research Institute of Thailand (Public Organization), Don Kaeo, Mae Rim, Chiang Mai 50180, Thailand}
\altaffiltext{13}{Harvard-Smithsonian Center for Astrophysics, 60 Garden St, Cambridge, MA 02138, USA}
\altaffiltext{14}{Department of Physics, Lehigh University, 16 Memorial Drive East, Bethlehem, PA 18015, USA}
\altaffiltext{15}{University of North Carolina at Chapel Hill}
\altaffiltext{16}{Event Horizon Telescope, Center for Astrophysics, Harvard \& Smithsonian, 60 Garden Street, Cambridge, MA 02138, USA}
\altaffiltext{17}{Vanderbilt University, Department of Physics \& Astronomy, 6301 Stevenson Center Ln., Nashville, TN 37235, USA }
\altaffiltext{18}{Fisk University, Department of Physics, 1000 17th Ave. N., Nashville, TN 37208, USA}

\begin{abstract}

The most dramatic phases of terrestrial planet formation are thought
to be oligarchic and chaotic growth, on timescales of up to 100--200
Myr, when violent impacts occur between large planetesimals of sizes
up to protoplanets. Such events are marked by the production of large
amounts of debris, as has been observed in some exceptionally bright
and young debris disks (termed extreme debris disks). Here we report
five years of {\it Spitzer} measurements of such systems around two
young solar-type stars: ID8 and P1121. The short-term (weekly to
monthly) and long-term (yearly) disk variability is consistent with
the aftermaths of large impacts involving large asteroid-sized
bodies. We demonstrate that an impact-produced clump of optically
thick dust, under the influence of the dynamical and viewing geometry
effects, can produce short-term modulation in the disk light
curves. The long-term disk flux variation is related to the
collisional evolution within the impact-produced fragments once
released into a circumstellar orbit. The time-variable behavior
observed in the P1121 system is consistent with a hypervelocity impact
prior to 2012 that produced vapor condensates as the dominant impact
product. Two distinct short-term modulations in the ID8 system suggest 
two violent impacts at different times and locations. Its
long-term variation is consistent with the collisional evolution of
two different populations of impact-produced debris dominated by
either vapor condensates or escaping boulders. The bright, variable
emission from the dust produced in large impacts from extreme debris
disks provides a unique opportunity to study violent events during the
era of terrestrial planet formation.

\end{abstract} 


\section{Introduction}

Planet formation is ubiquitous -- thousands of exoplanets have been
detected through Doppler spectroscopy, transit photometry,
microlensing surveys and direct imaging surveys, with each sensitive
to different populations of planets. However, our knowledge of the
formation process is generally limited to (1) the first $\sim$10 Myr:
studies of protoplanetary disks around young stars, and, recently, of
accretion onto forming giant planets
\citep{sallum15,johns-krull16,wagner18}; and (2) characterization of
the end results: planets orbiting mature stars \citep{winn18}.  The
situation is particularly daunting for studying terrestrial planet
formation, which extends well past the lifetime of protoplanetary
disks and produces exceedingly faint planets requiring currently
unobtainable high contrast and spatial resolution for their direct
detection.  Alternatively, transit observations are revealing mature
Earth-sized planets but provide little information about the
characteristics of their formation. Debris disks around mature stars
are excellent tools to search for phases occurring in other planetary
systems that are analogous to major events in the evolution of the
solar system, such as the formation of terrestrial planets
\citep{kenyon04,kenyon06} and the bombardment period in the early
solar system \citep{booth09,bottke17}.  Disk variability due to the
dust produced in the aftermaths of planetesimal impacts in young,
luminous debris disks provides a great opportunity to study the
violent events during the era of terrestrial planet formation
\citep{meng15,wyatt16}.

Models of terrestrial planet formation indicate that these rocky
planets grow via pair-wise accretion from planetesimal boulders
through runaway and oligarchic growth into planetary embryos, followed
by a final phase of giant impacts (e.g.,
\citealt{raymond14}). Numerical simulations suggest that this final phase
lasts for 100--200 Myr \citep{chambers13,genda15}. Assuming that the
impacts yield complete mergers in the $N$-body simulations, $\sim$10--15
giant impacts, defined as the collisions between two planetary
embryos, are required for the formation of an Earth-like planet
\citep{stewart12}. A significantly higher rate of smaller impacts
between embryos and asteroid-sized planetesimals is expected. Overall,
the impact rates would be higher if more realistic estimates of collisional
outcomes \citep{leinhardt12} were adopted \citep{chambers13}. The
diverse outcomes resulting from realistic collisions mean that the
impacts are less efficient to grow large bodies in general
\citep{agnor04}.  However, the frequency of impacts also increases
because the bodies resulting from the impacts that did not lead to net
growth (i.e., grazing and hit-and-run collisions) tend to come back
and collide with other bodies at a later time \citep{chambers13}. This
is why the timescale to build terrestrial planets remains similar to
the timescale with the perfect merger assumption.

Each giant impact is predicted to produce an observable signal due to
the production of huge clouds of dust and silica vapor
\citep{chambers98,kenyon06,jackson12,genda15,kenyon16}. Dust around
stars can be detected as an infrared excess, while its composition can
be studied through mid-infrared spectroscopy to reveal the presence of
debris material that went through shock and high-temperature events
(e.g., \citealt{morlok14}). About 1\% of the stars in the appropriate
age range for rocky planet formation have exceptionally large amounts
of warm circumstellar dust, indicative of high rates of collisional
activity that is expected to accompany active planet growth
\citep{balog09,melis10,kennedy13}. Because of their huge mid-infrared
excess emission above that of their stars (typical dust fractional
luminosity, $L_d/L_{\star}\gtrsim10^{-2}$), they are termed ``extreme
debris disks''.  The fraction of stars with huge infrared excesses
reaches $\sim$10\% in young ($\sim$25 Myr) clusters/associations
\citep{meng17}.

Interpreting these statistics in terms of overall terrestrial planet
formation models requires that we understand the individual systems,
including the duration of the observational signature of a major
impact (e.g., how rapidly the resulting infrared excess fades) and the
nature of the events we currently can observe and catalog. Thus,
characterization of these extreme systems to measure collisional
outcomes, both in terms of the unique composition of the products and
in their behavior in the time domain, can help reveal how
terrestrial planets grow. For the time domain, we have been using the
post-cryogenic {\it Spitzer} mission to monitor disk variability for a
dozen extreme debris disks in the past five years, with the main goals
to characterize the incidence, nature, and evolution of these
impacts. In this work, we report the results on ID8 (2MASS
J08090250$-$4858172) and P1121 (2MASS J07354269$-$1450422), two
solar-like stars that are known to possess a large infrared excess
accompanied by prominent 10 $\mu$m solid-state features, and that show
disk variability at [3.6] and [4.5] \citep{meng15}. The ages of both
stars coincide with the era of terrestrial planet formation (ID8 in
NGC2547 with an age of 35 Myr, and P1121 in M47 with an age of 80
Myr).

To observe an impact and its post-impact evolution, a frequent
cadence is needed. The frequency depends on the location of the dust,
which is within one au in both systems, as inferred from spectral
energy distribution (SED) modeling. The six-month cadence provided by
the {\it WISE} mission can only yield long-term information at most, which is 
inadequate to characterize any short-term variability. {\it Spitzer}
is the only available facility to do semi-regular infrared
monitoring. The wavelengths (3.6 and 4.5 $\mu$m) provided by the warm
{\it Spitzer} mostly trace material close in at small semimajor axes,
at which location the impact velocity can significantly exceed the
surface escape velocity of the impacting bodies. In our solar system,
Mercury is thought to have formed in a hypervelocity impact that
stripped the mantle material and left an anomalously large core
\citep{benz88,benz07}. We therefore expect evidence of similar violent
events in exoplanetary systems if an impact can be successfully
identified. Both ID8 and P1121 show such evidence.

The paper is organized as follows. Section \ref{obs} describes the
data and general results used in this work, including our warm
{\it Spitzer} data in Section \ref{spitzer_obs}, supplementary {\it
WISE} data in Section \ref{wise_obs}, and additional ground-based
optical monitoring and the resultant time-series disk fluxes and color
temperature trends in Section \ref{optical}. Detailed light curve
analyses in terms of short-term (weekly to monthly) and long-term
(yearly) behaviors are given in Sections \ref{analysis_id8} and
\ref{analysis_p1121} for the ID8 and P1121 systems, respectively.  We
also review and derive the general disk properties (dust location and
mass) using SED models, and discuss additional disk variability in the
mid-infrared wavelengths in Section \ref{debrislocation}. We then
interpret the observed variability due to the aftermath of an
impact-produced cloud of dust. The short-term semi-regular light-curve
modulations can be directly linked to the orbital evolution of an
optically thick cloud using a geometric and dynamical model developed
by \citet{jackson19}. We describe the basic idea of such a model,
derive the expected light curve modulations using 3D radiative
transfer calculations, and apply the results to the modulations in
both systems in Section \ref{interpretation_short-term}. We then focus
on the collisional evolution within the impact-produced cloud of dust
in Section \ref{interpretation_long-term} to qualitatively
explain the long-term disk variability.  A short discussion is given
in Section \ref{discussion}, followed by our conclusions in Section
\ref{conclusion}.

\section{Observations and Results  } 
\label{obs}

\subsection{{\it Spitzer} IRAC 3.6 and 4.5 \um Observations}
\label{spitzer_obs}

{\it Spitzer}/IRAC observations were obtained under GO programs PID
10157 (PI Rieke) and PID 11093, 13014 (PI Su). ID8 was monitored with
daily cadence under program PID 10157 from June to August 2014,
resulting in a total of 59 sets of observations in 2014. Both ID8 and
P1121 were monitored under PID 11093 and 13014 with $\sim$3-day
cadence during their visibility windows from 2015 to 2017, resulting
in a total of 220 sets of observations for ID8 and a total of 93 sets
of observations for P1121. For both objects, we used a frame time of
30 s with 10 cycling dithers (i.e., 10 frames per Astronomical
Observation Request (AOR)) to minimize the intrapixel sensitivity
variations of the detector \citep{reach05} at both [3.6] and [4.5],
achieving a signal-to-noise ratio (S/N) $>$100 in single-frame photometry.
These data were processed with the IRAC pipeline S19.2.0 by the {\it
Spitzer} Science Center.

Following the photometry procedure in \citet{meng15}, we performed
aperture photometry on the cBCD (Corrected Basic Calibrated Data)
images. An aperture radius of 3 pixels (3\farcs6) and an annulus of
12--20 pixels (14\farcs4--24\arcsec) were used with aperture
corrections of 1.112 and 1.113 at 3.6 \um and 4.5 \mm,
respectively. The cBCD photometry was also corrected for the pixel
solid angle (i.e., distortion) effect based on the measured target
positions. We obtained weighted-average photometry for each of the
AORs after throwing out the highest and lowest photometry points in the
same AOR. Finally, we used the median Barycenter Modified Julian Date
(BMJD) for each of the AORs as the time stamp for the weighted-average
photometry. The IRAC 3.6 and 4.5 \um data are not obtained
simultaneously, i.e., there is a typical time gap of 7.7 minutes between
the 3.6 and 4.5 \um observations.

\begin{deluxetable*}{ccrrrrcrrrr}
\tablewidth{0pc}
\footnotesize
\tablecaption{The IRAC fluxes of the ID8 system\label{obs_tab_id8}}

\tablehead{
\colhead{AOR Key}&\colhead{BMJD$_{3.6}$}&\colhead{$F_{3.6}$}&\colhead{$E_{3.6}$}&\colhead{$exeF_{3.6}$}&\colhead{$exeE_{3.6}$}  &\colhead{BMJD$_{4.5}$}& \colhead{$F_{4.5}$}&\colhead{$eF_{4.5}$}&\colhead{$exeF_{4.5}$}&\colhead{$exeE_{4.5}$}    \\ 
\colhead{  }&\colhead{(day) }&\colhead{(mJy)}&\colhead{(mJy)}&\colhead{(mJy)}&\colhead{(mJy)}&\colhead{(day)}&\colhead{(mJy)}&\colhead{(mJy)}&\colhead{(mJy)}&\colhead{(mJy)}
}
\startdata
  45677056 &    56072.71322 &     9.70 &     0.10 &     1.13 &     0.16 &    56072.70783 &     7.71 &     0.04 &     2.09 &     0.09  \\ 
  45677312 &        \nodata  &  \nodata &  \nodata &  \nodata &  \nodata &    56077.33125 &     7.58 &     0.03 &     1.95 &     0.09  \\ 
  45677568 &        \nodata  &  \nodata &  \nodata &  \nodata &  \nodata &    56087.51265 &     7.82 &     0.04 &     2.19 &     0.09  \\ 
  45677824 &        \nodata  &  \nodata &  \nodata &  \nodata &  \nodata &    56092.11433 &     7.70 &     0.04 &     2.07 &     0.09  \\ 
  45678080 &        \nodata  &  \nodata &  \nodata &  \nodata &  \nodata &    56099.02264 &     7.75 &     0.03 &     2.12 &     0.09  \\ 
  45678336 &        \nodata  &  \nodata &  \nodata &  \nodata &  \nodata &    56108.67032 &     7.80 &     0.02 &     2.17 &     0.09  
\enddata
\tablecomments{$F$ and $E$ are the flux and uncertainty including the
star, while $exeF$ and $exeE$ are the excess quantities excluding the
star.  Table \ref{obs_tab_id8} is published in its entirety in the
machine-readable format. A portion is shown here for guidance
regarding its form and content.}
\end{deluxetable*}

\begin{deluxetable*}{ccrrrrcrrrr}
\tablewidth{0pc}
\footnotesize
\tablecaption{The IRAC fluxes of the P1121 system\label{obs_tab_p1121}}
\tablehead{
\colhead{AOR Key}&\colhead{BMJD$_{3.6}$}&\colhead{$F_{3.6}$}&\colhead{$E_{3.6}$}&\colhead{$exeF_{3.6}$}&\colhead{$exeE_{3.6}$}  &\colhead{BMJD$_{4.5}$}& \colhead{$F_{4.5}$}&\colhead{$eF_{4.5}$}&\colhead{$exeF_{4.5}$}&\colhead{$exeE_{4.5}$}    \\ 
\colhead{  }&\colhead(day)&\colhead{(mJy)}&\colhead{(mJy)}&\colhead{(mJy)}&\colhead{(mJy)}&\colhead{(day)}&\colhead{(mJy)}&\colhead{(mJy)}&\colhead{(mJy)}&\colhead{(mJy)}
}
\startdata
  45680640 &    56077.83283 &    11.74 &     0.06 &     2.25 &     0.15 &    56077.82722 &     9.18 &     0.02 &     3.01 &     0.10  \\ 
  48054272 &    56311.19693 &    11.38 &     0.06 &     1.89 &     0.16 &    56311.19135 &     8.79 &     0.03 &     2.62 &     0.10  \\ 
  48054528 &    56315.44559 &    11.07 &     0.06 &     1.58 &     0.16 &    56315.44000 &     8.44 &     0.03 &     2.27 &     0.10  \\ 
  48054784 &    56318.87263 &    11.08 &     0.05 &     1.59 &     0.15 &    56318.86703 &     8.51 &     0.02 &     2.34 &     0.09  \\ 
  48055040 &    56323.93743 &    11.03 &     0.03 &     1.54 &     0.15 &    56323.93180 &     8.42 &     0.02 &     2.25 &     0.09  \\ 
  48055296 &    56329.12567 &    11.23 &     0.08 &     1.75 &     0.16 &    56329.12001 &     8.57 &     0.02 &     2.40 &     0.09  
\enddata
\tablecomments{$F$ and $E$ are the flux and uncertainty including the
star, while $exeF$ and $exeE$ are the excess quantities excluding the
star. Table \ref{obs_tab_p1121} is published in its entirety in the machine-readable format. A portion is shown here for guidance regarding its form and conten.}
\end{deluxetable*}

To evaluate the uncertainty and stability of the IRAC photometry, we
selected a handful of stars in the field of view as references, and
obtained their photometry as described above. These reference stars
have similar or fainter fluxes than our targets, and the measured
stability is within 1.2\% at both wavelengths, consistent with the
expected repeatability of the instrument \citep{rebull14} over
multiyear timescales. Based on the repeatability of the
reference stars, we conclude that photometry variation above 3\%
levels is significant and has an astrophysical origin.

\subsection{WISE photometry }
\label{wise_obs}

We extracted {\it WISE} 3.4 $\mu$m ($W1$) and 4.6 $\mu$m ($W2$) photometry from
the {\it WISE} \citep{wright10} and NEOWISE
\citep{mainzer11,mainzer14} missions through the IRSA archive
maintained by IPAC. Because we are interested in the time-domain
photometry, we searched the single-exposure source table by matching the
target position within 10\arcsec\ in the four major {\it WISE}
surveys\footnote{{\it WISE} Cryogenic Survey, {\it WISE} 3-band Survey, {\it WISE}
Post-Cryo Survey, and {\it WISE} Reactivation, details see
http://irsa.ipac.caltech.edu/Missions/wise.html} that cover the {\it WISE}
data up to September 2016. All single-frame photometry was time-averaged 
to match the cadence of {\it Spitzer} monitoring ($\sim$3
days). The {\it WISE} magnitudes were then transferred to flux density
units by adopting the zero-point fluxes from \citet{wright10} and
\cite{jarrett11}. Because the filters are not identical between {\it
Spitzer} and {\it WISE}, we applied a uniform flux offset per band in
comparing the {\it WISE} photometry to the {\it Spitzer} data. These
{\it WISE} points are used to assess the long-term trend of the disk
variability, especially during the gaps between the {\it Spitzer}
visibility windows.  The bulk of the analysis (Section 3) is based on
the {\it Spitzer} observations.

\subsection{Time-series Excess Fluxes and Color Temperatures} 
\label{optical}

The stars in both systems have been intensively monitored from the
ground in the optical $V_c$ ($\lambda_{eff}$=0.54 $\mu$m) and $R_c$
($\lambda_{eff}$=0.64 $\mu$m) bands during 2013 \citep{meng14,meng15},
and the optical fluxes were found to be stable within 1\%. During the
{\it Spitzer} visibility windows, we continued to monitor both systems
using the 0.41 m PROMPT8 robotic telescope at Cerro Tololo
Inter-American Observatory in Chile whenever the conditions
permitted. Both stars are again stable within 1--2\% levels. We also
obtained additional optical data from the KELT network
\citep{pepper07, pepper12} and the ASAS-SN project
\citep{shappee14,kochanek17}. For ID8, there were 1335 observations
collected by KELT from 2012 September to 2014 April using a
nontraditional broad-R filter and with a typical error of 0.04
mag. There were 500 observations available from the ASAS-SN project
from 2016 February to 2018 March with a typical error of 0.02 mag. We
searched for periodicity in these optical data using the SigSpec code
\citep{reegen07}, and found a period of 5$\pm$1 days with an amplitude
of 0.013 mag in the ASAS-SN data. This confirms the previous result
from \citet{meng14}, where the weak (0.01 mag) 5-day modulation is
attributed to spots on the stellar surface, showing that the rotation
axis of ID8 is unlikely to be pole-on from our line of sight.  For
P1121, there were $\sim$1600 observations from KELT (spanning from
2013 May to 2017 October with a typical error of 0.04 mag), and
$\sim$800 observations from ASAS-SN (spanning from 2012 Jan to 2018
March with a typical error of 0.02 mag). No significant periodicity of
more than 1 day was found for P1121. Finally, no detectable optical
eclipse was found in all available optical data, suggesting that the
orientation of both systems is not likely to be exactly edge-on, unlike
the RZ Psc system \citep{kennedy17}, one of the extreme debris disks
that show infrared variability (K.\ Su et al.\ in preparation).

Given the stability of the stellar output, we obtained the disk fluxes
by subtracting the expected photospheric fluxes at each band. We first
evaluated the photospheric values predicted by Kurucz atmospheric
models in light of the distance given by the {\it Gaia} DR2 catalog
(361$\pm$2 pc for ID8 and 459$\pm$7 pc for P1121,
\citealt{gaia16,gaia18}). The ID8 photospheric fluxes (8.56 mJy and
5.63 mJy at the [3.6] and [4.5] bands, respectively) given by
\citet{meng14,meng15} are consistent with the values from a
main-sequence dwarf ($L_{\ast}= 0.72 L_{\sun}$) with a spectral type
of G6 at 360 pc and a modest (0.03 mag) interstellar extinction. For
P1121, the {\it Gaia} DR2 catalog gives a stellar effective
temperature of 5856 K, which is slightly lower than the 6200 K used by
\citet{meng15}. Using the {\it Gaia} temperature, we derived the
photospheric fluxes of 9.49 and 6.17 mJy at the [3.6] and [4.5]
bands, consistent with a G0 dwarf ($L_{\ast}= 1.48 L_{\sun}$) at 459
pc and with an interstellar extinction of 0.2 mag. This type is
consistent with the spectroscopic classification of F9 IV--V
\citep{gorlova04}. We note that the newly adjusted photospheric fluxes
are still within the 2\% uncertainty of the previously estimated
values.

At [3.6] and [4.5], the stellar photosphere contributes more than 50\%
of the total output at both bands; therefore, the uncertainty for the
estimated disk flux (excess) is dominated by the star. The estimated
disk flux uncertainty includes typical errors of 1.5\% from the
photospheric extrapolation and the nominal photometry uncertainty from
the weighted average. For consistency, we also remeasured the
photometry using the data published in \citet{meng14,meng15}.  The
final time-series measurements are given in Table \ref{obs_tab_id8}
for ID8 and Table \ref{obs_tab_p1121} for P1121.

We computed the color temperatures of the excesses by ratioing the
disk fluxes at both bands. Given the small wavelength difference
between the two IRAC bands, the color temperatures are only an
indication of the dust temperatures in a relative sense to monitor the
overall trend. However, the emission we detected is most likely to be
a combination of optically thick and thin emission (as discussed in
Sections \ref{interpretation_short-term} and
\ref{interpretation_long-term}); inferring dust location from such
disk color temperatures is rather complicated.  Furthermore, the star
dominates the noise in the measured excess; therefore, the derived
color temperatures inherit these uncertainties, resulting in a typical
error of $\sim$100 K in the individual color temperatures.  To better
illustrate the overall trend, time-averaged (one to a few per
visibility window) color temperatures are also derived. Figure \ref{timeseries_id8_p1121}
shows the time-series disk fluxes and the
corresponding color temperatures for the ID8 and P1121 systems.

\begin{figure*} 
  \figurenum{2}
  \label{id8_2014} 
 \epsscale{1.15} 
  \plotone{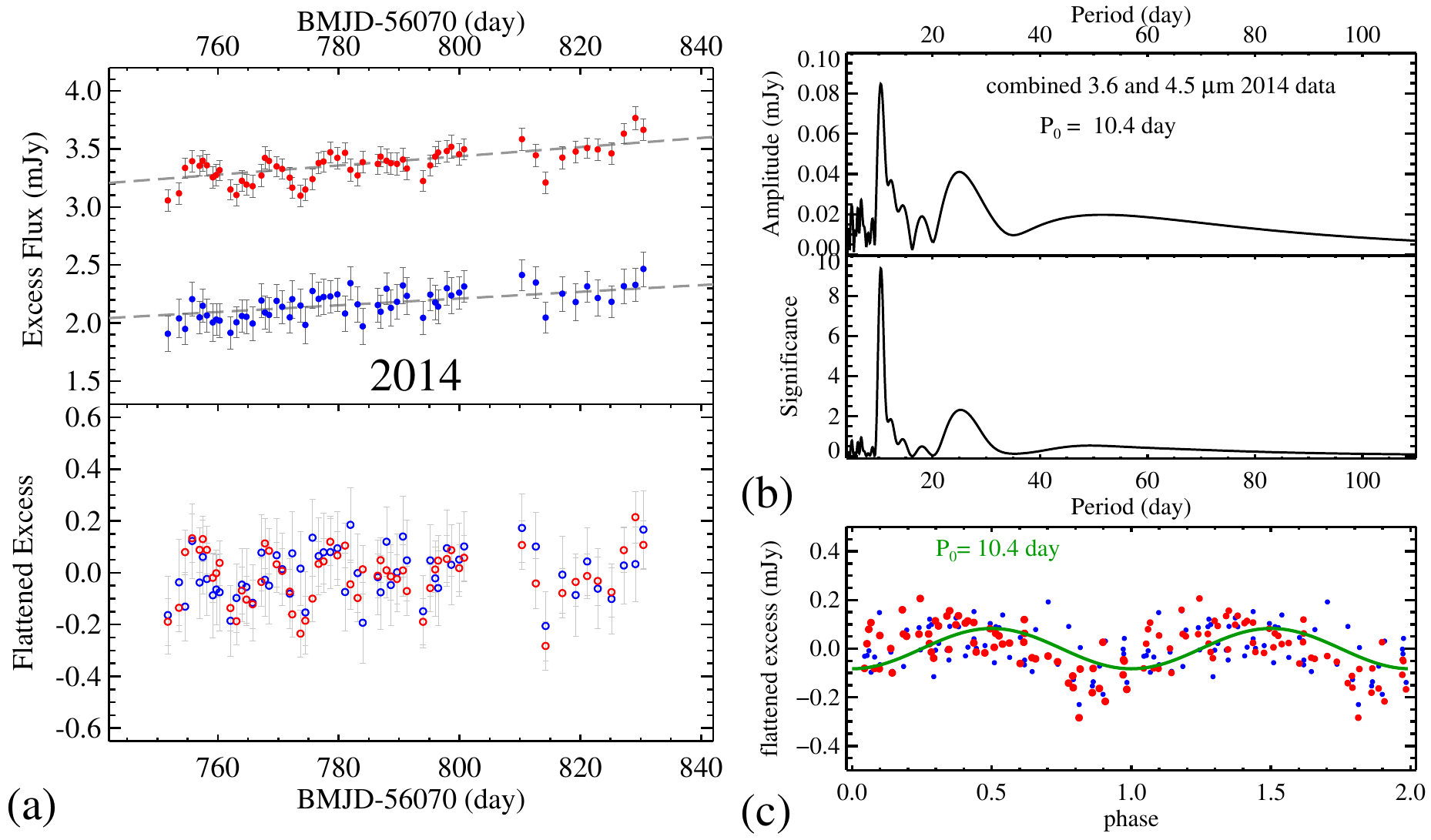}
  \caption{(a) The top panel shows the ID8 disk fluxes observed in
2014 (roughly daily cadence for $\sim$80 days). The red color
represents the 4.5 $\mu$m band, while the blue is for the 3.6 $\mu$m
band. The dashed lines are the derived general trends. The bottom
panel shows the flattened excess after subtraction of the fitted
linear trend (see Sec 3.1). (b) Periodicity analysis for the 2014
ID8 flattened excess using the SigSpec code. The top and bottom panels
show the associated amplitude and significance (i.e., S/N) of the
period. A significant and sharp peak is found at 10.4 days. (c) The
folded phase curve of the 2014 data (dots) with fitted sine curve
(green line).}
\end{figure*}

\begin{figure*} 
  \figurenum{3}
  \label{id8_2013567} 
 \epsscale{1.15} 
  \plotone{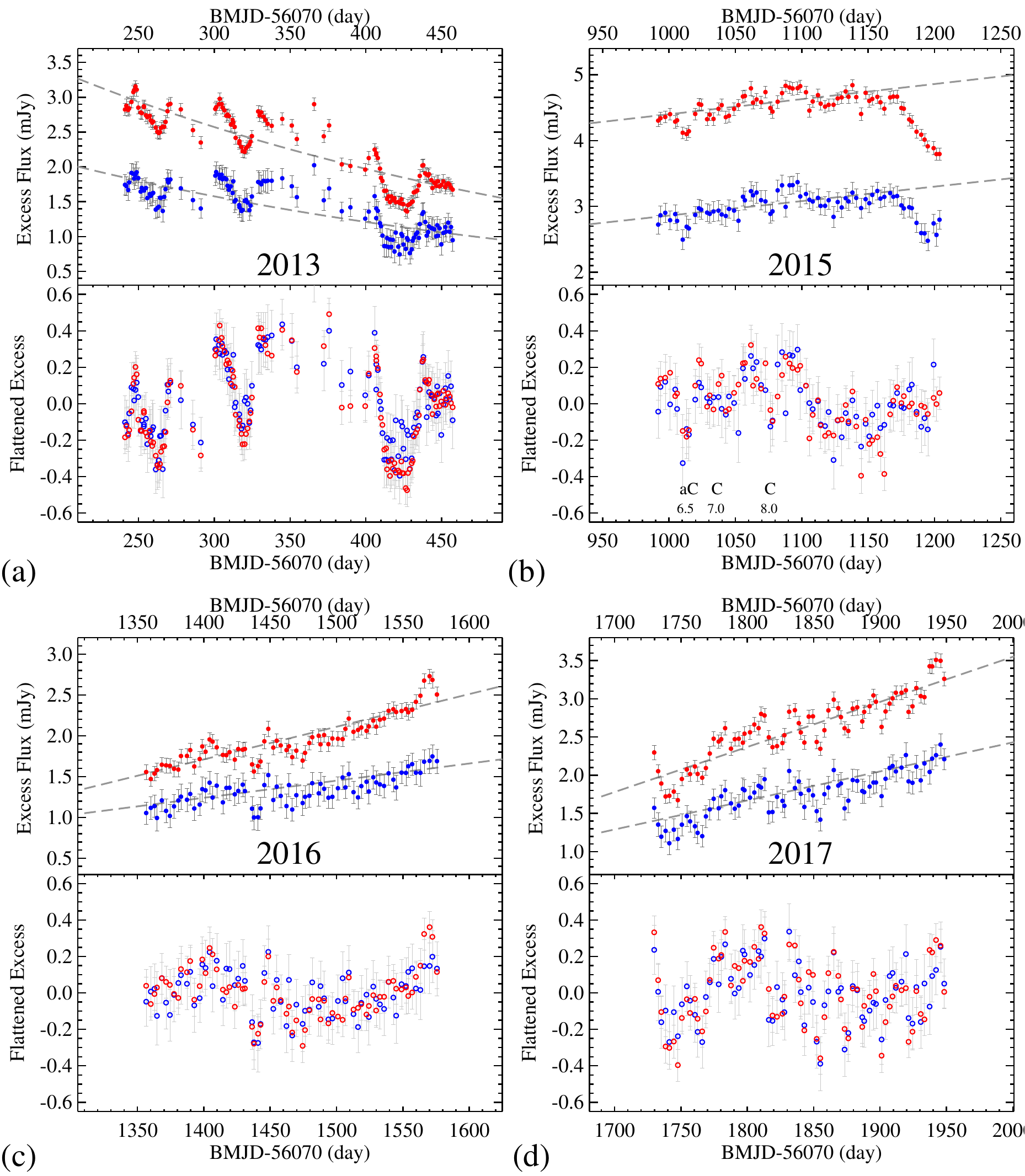}
  \caption{ID8 disk fluxes and flattened excesses for data taken in
2013 (a), 2015 (b), 2016 (c), and 2017 (d). The symbols used are the
same as in Figure \ref{id8_2014}(a). In all panels, the x-axis shows
the same range of 280 days. The range of the y-axis in the upper panel
(disk flux) is different from year to year, but the bottom (flattened
excess) panel has the same range. In the bottom panel of the 2015 disk
light curve, we also mark the possible collision (C) and
anti-collision (aC) dips and associated orbital phases (numbers) due
to the 2014 impact events (details see Section \ref{modulations}). }
\end{figure*}

\begin{figure*} 
  \figurenum{4}
  \label{p1121_decay} 
 \epsscale{1.15} 
  \plotone{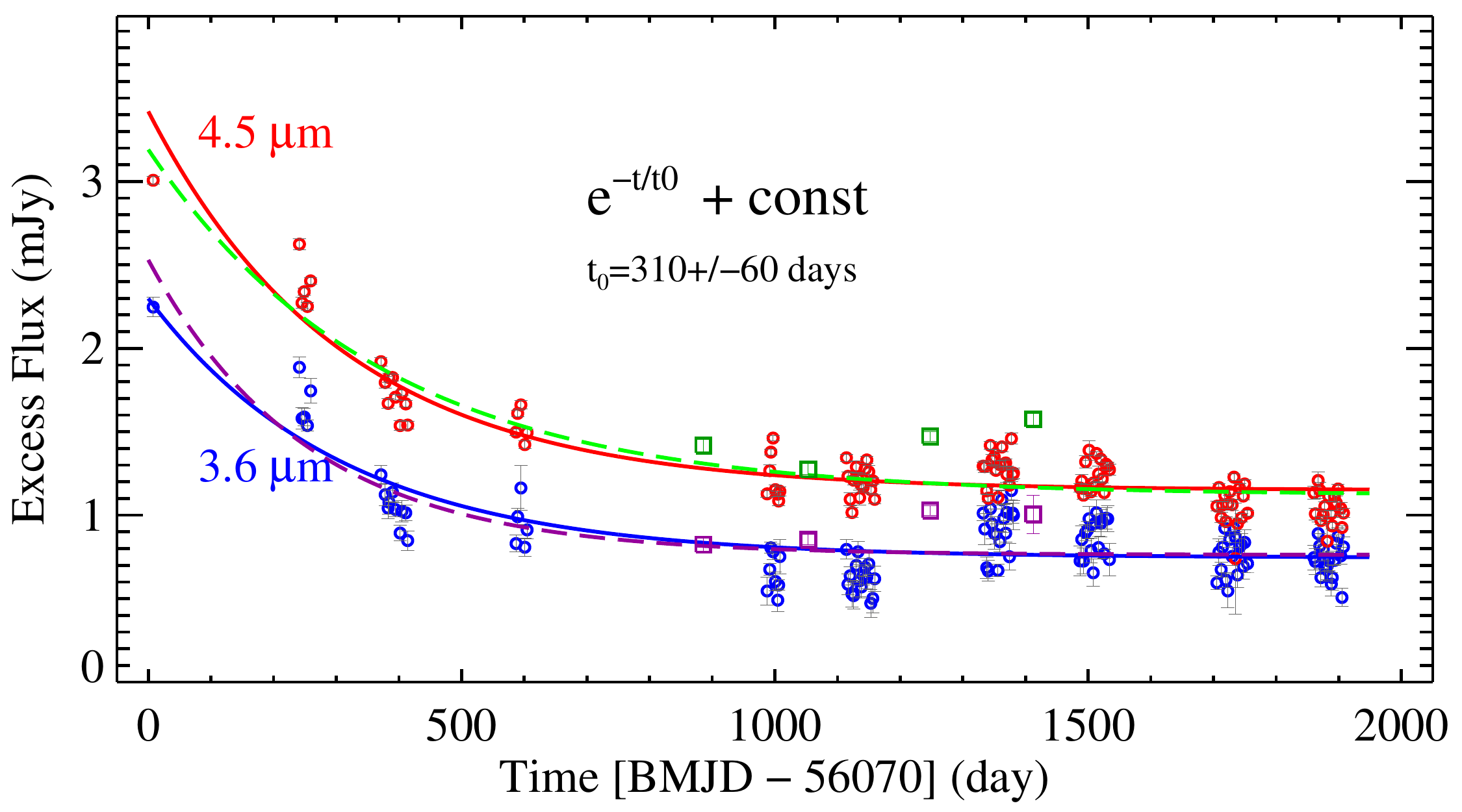}
  \caption{Decay fits to the P1121 data. Solid lines (red or blue)
are the fits with the same decay timescale ($t_0$) of$\sim$ 310 days. 
Dashed lines (green or purple) are the fits without fixing the decay-time 
constant; in this case, $t_0$ is found to be $\sim$253 days at
3.6 $\mu$m and $t_0$ is found to be $\sim$370 days at 4.5 $\mu$m.}
\end{figure*}

\begin{figure*} 
  \figurenum{5}
  \label{p1121_fft} 
 \epsscale{1.15} 
  \plotone{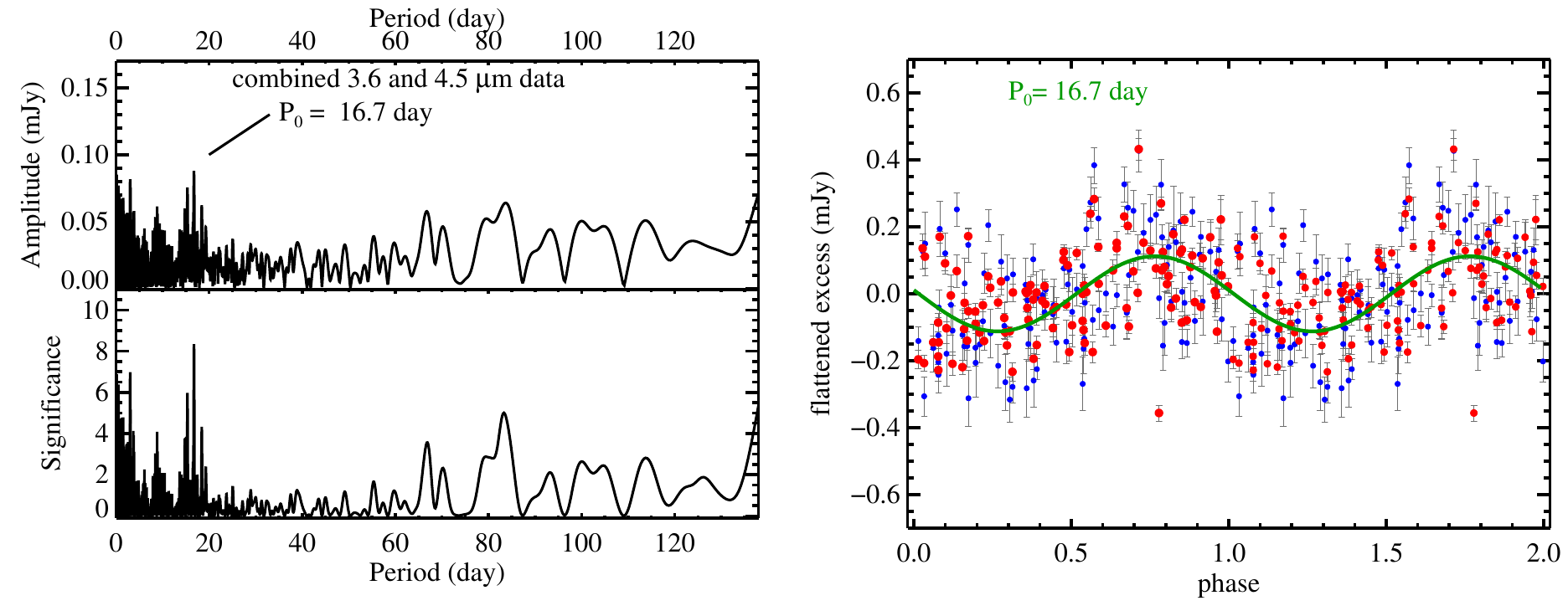}
  \caption{(a) Periodicity analysis (SigSpec code) of the P1121 data
since 2012 after the general decline trend is subtracted. The top
and bottom panels show the associated amplitude and significance of
the period. (b) The folded phase curve using the period of 16.7
days. }
\end{figure*}

\section{Analysis:  Temporal Behavior and General Disk Properties }
\label{analysis}

\subsection{ID8}
\label{analysis_id8}

\begin{deluxetable*}{cccccccccc}
\tablewidth{0pc}
\tablecaption{Linear slopes\tablenotemark{a} of the Increase/Decrease in the ID8 disk flux\label{id8_slopes}}
\tablehead{
\colhead{Band} & \colhead{2012} & \colhead{2013} & \colhead{2014} & \colhead{2015\tablenotemark{b}} & \colhead{2015\tablenotemark{c}} & \colhead{2014/2015\tablenotemark{d}} & \colhead{2016} & \colhead{2017} & \colhead{2016/2017}
}
\startdata
$[3.6]$  &\nodata      &$-$3.8$\pm$0.1 & 3.9$\pm$0.9 & 2.2$\pm$0.4 &$-$20.8$\pm$4.0 & 2.9$\pm$0.1 & 2.1$\pm$0.3 & 3.8$\pm$0.3 & 1.3$\pm$0.1 \\
$[4.5]$  & 1.4$\pm$1.0 &$-$6.3$\pm$0.2 & 4.4$\pm$0.5 & 2.3$\pm$0.2 &$-$26.3$\pm$2.1 & 4.0$\pm$0.1 & 4.0$\pm$0.2 & 5.9$\pm$0.2 & 2.1$\pm$0.1 \\  
\enddata
\tablenotetext{a}{in units of $\mu$Jy day$^{-1}$.}
\tablenotetext{b}{the first part of the 2015 data where fluxes increase.}
\tablenotetext{c}{the second part of the 2015 data where fluxes decrease. The last three points at 3.6 $\mu$m were excluded from the fit.}
\tablenotetext{d}{combining the 2014 and the first part of the 2015 data where fluxes increase.}
\end{deluxetable*}

Similar to the variability observed in 2013 \citep{meng14}, the disk
fluxes at both bands track each other relatively well. Unlike the disk
flux decay observed in 2013, most of the excesses in the new {\it
Spitzer} observations showed an upward trend, except for the short
($\sim$50 days) period near the end of the 2015 window (see Figure
\ref{timeseries_id8_p1121}a). The upward trend appeared to start as
early as the end of the 2013 light curves. The steep decline near the
end of the 2015 appeared to continue until the beginning of 2016. The
{\it WISE} point near the displayed day\footnote{Hereafter, we use
``d.d.'' as the displayed day in the text that references BMJD 56070
as the zero-point.} of 1300 (d.d.\ 1300) corroborates this rapid
decline.  In the past five years of {\it Spitzer} monitoring, the disk
flux reached the lowest value near the end of 2013 at $\sim$8\% and
$\sim$20\% excesses above the photosphere at 3.6 and 4.5 \mm,
respectively, and the highest in mid-2015 at $\sim$40\% and
$\sim$87\%, respectively. The average color temperature of the disk
over 5 yr is 731 K, with a 1 $\sigma$ standard deviation of 50 K.
Overall, there is no significant trend between the disk flux and the
observed color temperature.

\citet{meng14} found short-term variations associated with two
intermixed periodicities. Semi-regular up-and-down patterns on top of
the long-term trends are also seen in the 2014 and 2015 data. Before
searching for periodicities that might fit the data, we first
determined the overall trends for the new 2014--2017 observations.  To
minimize the free parameters, we fit a linear function to various
segments of the data. The fitted slopes (in units of $\mu$Jy day$^{-1}$) are
listed in Table \ref{id8_slopes}. Generally, the increasing rates are
very similar at [4.5]. We
also determined the linear slope for the 2013 data (instead of an
exponential decay as described in \citet{meng14}). The decline in disk
fluxes near the end of 2015 is very rapid, $\gtrsim$4 times faster
than in 2013.  We will discuss the implications of the long-term
upward and downward trends in Section \ref{interpretation_long-term}.

After the general flux trends were removed, we used the SigSpec code
to search for periodicity in the ``flattened'' excesses. Various
different combinations of data segments were searched either per band
or combining both bands. In the new 2014--2017 observations, only the
2014 data show an obvious periodicity, 10.4$\pm$1.0 days, as shown in
Figure \ref{id8_2014}. This period is much shorter than the two
periods found in the 2013 data ($\sim$26 and $\sim$33 days). The
modulation amplitude ($\pm$0.08 mJy, see Figure \ref{id8_2014}) is
similar to 2013 ($\pm$0.16 mJy for the 33-day period, and $\pm$0.08
mJy for the 26-day period, see Figure \ref{id8_2013567}a). Because the
visibility window for ID8 is about $\sim$220 days long each year, any period
longer than $\sim$110 days found in one-year data is not considered
significant. For reference, the segments of the disk fluxes and
flattened excesses in 2013 and 2015--2017 are also shown in Figure
\ref{id8_2013567}. When the whole 5 yr of data are combined for the
Fourier analysis, several long periods also appear: $\sim$148,
$\sim$184, and $\sim$360 days. We considered these periods as aliases
due to sampling effects because the associated peaks in the
periodogram are broad, and we also obtained similar periodicity using
the photometry of the reference stars that is stable within 2\% in the
IRAC photometry. In summary, the semi-periodic behavior is only
found in the data segments of 2013 and 2014 with very different
periodicities between the two (two intermixed periods in 2013, but a
different single period in 2014). The single 2014 periodicity appeared
to persist until mid-2015 and had no trace afterward, suggesting that
whatever caused the short-term modulation also needs to be less
effective as time goes on. The disappearing nature is an important
clue for understanding the cause of the short-term modulations (see
Section \ref{interpretation_short-term} for further discussion).

\subsection{P1121} 
\label{analysis_p1121}

Similar to ID8, the disk fluxes at both bands track each other pretty
well. Unlike ID8, the overall disk flux in the P1121 system appears
to be relatively quiescent since 2015. Using the {\it WISE} data to
fill the gaps between {\it Spitzer} windows, the disk flux in the 3--5
$\mu$m range appeared to be the highest in 2012, then followed a
general decline to the 2015/2017 quiescent level. To quantify the
decay rate, we fit an exponential plus a constant function
($e^{-t/t_0}+ C$) to the data obtained since 2012. Both 3.6 and 4.5
$\mu$m data can be well fit with the same decay timescale,
$t_0$=310$\pm$60 days (Figure \ref{p1121_decay}). This decay constant
is quite similar to the one found in the 2013 disk flux in the ID8
system, i.e., on the order of one year \citep{meng14}.  The quiescent
disk flux (background disk emission) is 0.77 mJy at [3.6] and 1.16 mJy
at [4.5], suggesting a color temperature of $\sim$750 K.  The average
color temperature over 5 yr is 751 K, with a 1 $\sigma$ deviation of
147 K (Figure \ref{timeseries_id8_p1121}b). It appears that the color
temperature decreases as the disk flux decreases from 2012 to
2015. The average color temperature in late 2012 and early 2013 is
$\sim$800$\pm$32 K, while the average color temperature in 2015 is
610$\pm$60 K. The average color temperature in 2017 (820$\pm$200 K) is
slightly higher than the average 2012/2013 value ($\sim$800 K) when
the disk flux is the highest.

On top of this general flux decline, a small modulation is also
seen. We performed a Fourier analysis (SigSpec code) on the combined
flattened light curves (shown in Figure \ref{p1121_fft}). Two periods
(647$\pm$3 days and 16.7$\pm$1.5 days) have significance (i.e., S/N)
above 8. When only using the 4.5 $\mu$m data, the 647-day period
disappears; however, the 16.7-days period persists in either single or
combined 3.6 and 4.5 $\mu$m data. Because we only have sparse data
points for a period of $\sim$1900 days and the long-term periodicity
also depends sensitively on the assumed function of the general flux
trend, a periodicity shorter than $\sim$3 days (monitoring cadence) and
longer than $\sim$300 days cannot be well constrained with the current
data.  We also searched for periodicity using the photometry of the
reference stars, and no period shorter than $\sim$300 days was
present. At this point, we only consider the period of 16.7 days to be
genuine. The bottom panel of Figure \ref{p1121_fft} shows the folded
disk phase curves at both bands. Interestingly, the modulation
amplitude ($\pm$0.08 mJy) is very similar to the one found in the 2014
periodicity in the ID8 system (Figure \ref{id8_2014}).

\begin{figure*} 
  \figurenum{6}
  \label{id8_p1121sed} 
  \epsscale{1.15} 
  \plotone{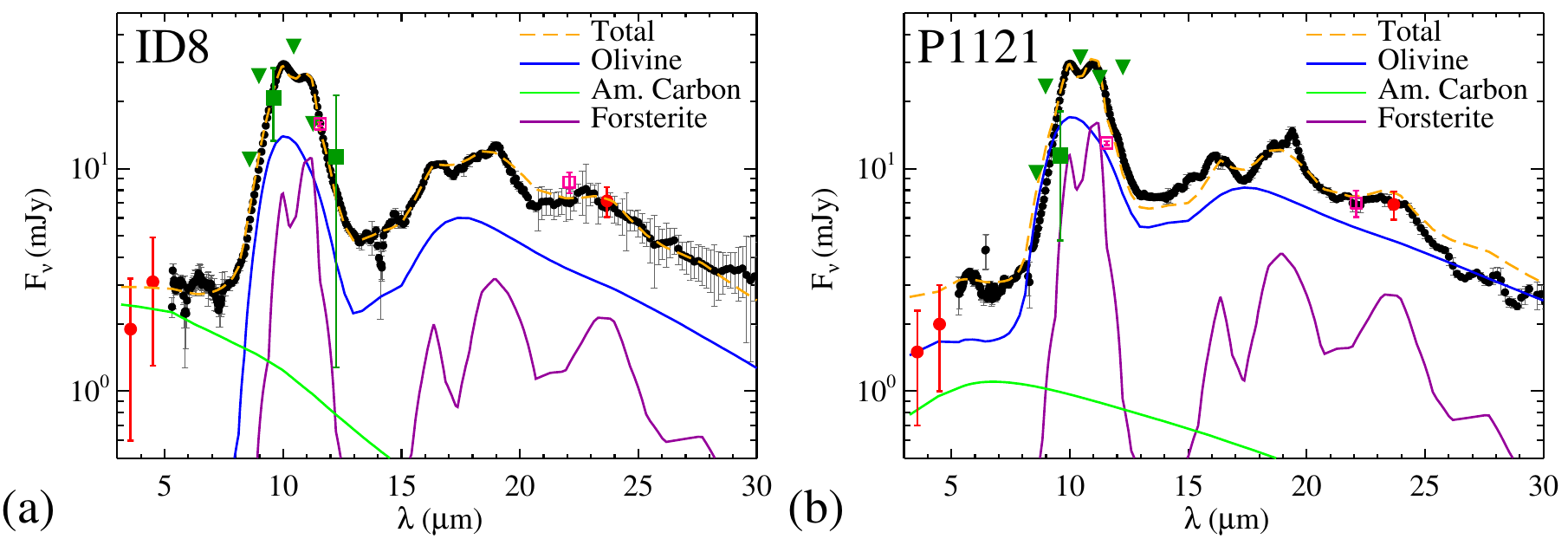}
  \caption{{\it Spitzer} IRS spectra (black dots) of the ID8 (a) and
P1121 (b) debris systems.  The stellar contribution has been
subtracted in both panels. Various solid lines are the
contributions of three major dust compositions (olivine: magnesium
iron silicate, forsterite: magnesium-rich end member of olivine, and
amorphous carbon) used in the SED model developed by
\citet{olofsson12} (details see Section \ref{debrislocation}). The red
dots are the {\it Spitzer} measurements, with error bars showing the
range of flux variation. The ALLWISE $W3$ and $W4$ points are shown
as the pink squares (taken in 2010). The dark green squares and
downward triangles are the VLT/VISIR fluxes and 3$\sigma$ upper limits
(taken in 2015). }
\end{figure*}

\subsection{Debris Location Inferred from SED Models}
\label{debrislocation}

To have a complete view of the two systems and properly interpret the
causes of variability at [3.6] and [4.5], we first sketch the general
disk properties (dust location and mass) using SED models, and discuss
additional disk variability in the mid-infrared wavelengths. Both ID8
and P1121 show prominent solid-state features in their mid-infrared
spectra, suggesting the presence of abundant small silicate-like
grains in the system. \citet{olofsson12} presented a detailed study
for the ID8 system by simultaneously determining dust composition and
disk properties. They found that $\sim$10\% of the small dust in the
ID8 system is in the form of crystalline silicates with about two-thirds of
them belonging to Fe-rich crystalline grains, in stark contrast to the
crystalline silicates found in the gas-rich protoplanetary disks where
Fe-bearing crystalline grains are rarely observed. In their model, the
spatial distribution of the dust is assumed to be an optically thin,
flat disk described by the parameters of inner and outer radii ($r_{in}$
and $r_{out}$), and a surface density power-law index ($\alpha$). The
prominent features suggest that the grains in the disk are dominated
by submicron sizes in a steep power-law size distribution (a
power-law index, $p$, of --4) with a total dust mass of
2.4$\times10^{-6}$ M$_{\earth}$ (up to 1 mm). The location of the
debris is estimated to be 0.32--0.64 au with $\alpha = -2.2\pm0.9$,
i.e., heavily peaked at the inner radius \citep{olofsson12}. We show
one of their best-fit model SEDs in Figure \ref{id8_p1121sed}a for
reference.

For consistency, we used the same approach and obtained a similar SED
model for the P1121 system to derive the model parameters as in
\citet{olofsson12}. We used the archival {\it Spitzer} IRS spectrum
from the CASSIS website that provides uniform, high-quality IRS
spectra optimally extracted for point-like sources
\citep{lebouteiller11}. One of the best-fit model SEDs is shown in
Figure \ref{id8_p1121sed}b. Compared to the ID8 system, the fraction
of crystalline silicate grains is higher, $\sim$40\%, although the
fraction of Fe-rich crystalline grains is similar, about two-thirds. 
The power-law index, $p$, in the grain size distribution is
$-$3.34$\pm$0.04. The location of the debris ranges from 0.2 to 1.6 au
with $\alpha=-1^{+0.1}_{-3.2}$, again favoring a close-in
location. The total dust mass in the P1121 system is
9.0$\times10^{-6}$ M$_{\earth}$ (up to 1 mm size). In this mineralogy-driven 
model, the emission at the two IRAC wavelengths mostly comes
from the amorphous carbon grains. However, in reality, it is difficult
to confirm their presence due to the lack of strong features in the
mid-infrared. This featureless disk emission can also come from the
contribution of large grains in the system whose mass contribution is
not captured in our derived dust mass. We also note that there is
no sign of small ($\sim$$\mu$m) silica grains present in both systems
when the mid-infrared spectra were taken due to lack of the distinct 9
$\mu$m feature. However, we cannot rule out the presence of large
silica grains.

It is interesting to note that crystalline grains in both systems are
dominated by the Fe-rich silicates, similar to other warm debris disks
modeled by \citet{olofsson12}. \citet{morlok14} present a detailed
mineralogical comparison between the dust composition in extreme
debris disks and that of meteorites, and suggest that the material
(which can be directly traced by the disk SED) in both ID8 and P1121
is similar to the material produced in high-temperature events with
relatively weak shocks (see their Figure 4).

Both IRS observations of the two systems were obtained in 2007 (ID8 in
2007 Jun 16 and P1121 in 2007 Apr 25), five years earlier than the
start of our {\it Spitzer} monitoring. The disk variability in ID8 was
first discovered by \citet{meng12} with a 10--30\% peak-to-peak
variation at 24 $\mu$m using {\it Spitzer} data obtained from 2003 to
2007. For P1121, we also computed synthesized 24 $\mu$m photometry by
integrating the 2007 IRS spectrum with the bandpass, which gives a
flux density of 6.4$\pm$0.3 mJy. Compared to the 2003 MIPS 24 $\mu$m
measurement from \citet{gorlova04}, the 24 $\mu$m flux dropped by
$\gtrsim$10\% over a few years. To test whether the photometric
variation seen by {\it Spitzer} is accompanied by spectral variation,
which might arise from changes in the dust size distribution, for
example, we observed ID8 and P1121 with the VLT/VISIR instrument in
late 2015 using six narrowband filters near 10 $\mu$m (PI: Kennedy,
ID: 095.C-0759(D)). These data were processed using the ESO pipeline
and corrections for calibrators observed at different airmasses using
the method outlined by \citet{verhoeff12}. Both targets were detected
(S/N $\gtrsim$2) in the J9.8 filter ($\lambda_{eff}$=9.6 $\mu$m and
$\Delta \lambda$=1 $\mu$m), the widest filter of the six
filters. The VISIR fluxes and 3 $\sigma$ upper limits are shown in
Figure \ref{id8_p1121sed}. The 2010 ALLWISE $W3/W4$ points are
also shown in Figure \ref{id8_p1121sed}, corroborating that there is
no dramatic change in the solid-state features. Overall, the
ground-based 10 $\mu$m observations were not sensitive enough to place
strong constraints on the spectral variation.  However, the VLT/VISIR
data suggest that some amount of small grains persists over $\sim$7--8
yr.

Given the degenerate nature between the grain properties and disk
location in the SED modeling, the exact distribution of the debris
cannot be well constrained from the SED model, i.e., a narrow-ring
peaked at 0.2, 0.3, or 0.5 au with slightly different grain properties
can also give satisfactory fits to the observed spectrum. Furthermore,
both systems lack data longward of 30 $\mu$m, therefore we cannot
rule out a faint, outer ($>$5 au) disk component either. We stress
that the calculations in the SED models assume that the dust is
optically thin (a low-density region where optical depth is much lower
than 1), which is a legitimate assumption for the strong solid-state
features. The observed disk SEDs are likely a mixture of optically
thin and thick components, as we discuss in Section
\ref{interpretation_short-term}.  Given their variable nature, the
debris location derived from one single epoch of the mid-infrared
spectrum should be taken with caution. It is possible that all or
most of the variations seen in [3.6] and [4.5] comes from dust closer
to the star and is separated from the dust that accounts for most of
the mid-infrared emission.

\section{Interpretation: Short-term Modulation}
\label{interpretation_short-term}

Debris generated by a violent impact forms a thick cloud of fragments.
As the impact-generated fragments are further dynamically sheared by
the Keplerian motion as they orbit the star, they also
collide among themselves to generate fine dust that emits efficiently
in the infrared. We posit that the complex infrared variability in
both systems can be explained by the combination of the dynamical and
collisional evolution from an impact-produced cloud.  Given the large
range of particle sizes involved in such an impact-produced cloud, it
is numerically challenging to couple the dynamical and collisional
evolution of the cloud self-consistently (e.g., \citealt{kral15}). We
therefore qualitatively model the short-term and long-term
variability separately using existing codes to extract basic
parameters of the impacts.

We first focus on the interpretation of the short-term disk flux
modulations, which can be explained using a geometric and dynamical
model from an optically thick cloud of dust produced in a violent
impact \citep{jackson19}. We describe the basic idea of the model in
Section \ref{basicideas}, and verify the expected disk flux modulation
using 3D radiative transfer calculations in Section \ref{3drt}. In
Section \ref{modulations} and \ref{modulations_p1121}, we apply our
model to the modulations seen in ID8 and P1121, and derive the impact 
locations and the likely impact dates.

\subsection{Basic Ideas}
\label{basicideas}

It is beneficial to first review the previous ideas in explaining the
temporal behavior in the 2012/2013 ID8 disk light curve, and establish
some basic parameters in these two systems. The amount of IRAC [3.6]
and [4.5] excess flux and variability around both systems cannot come
from a rigid body around the star given the known distance. At the
distance ($d$) of 360 pc for ID8, an object with a one-Jupiter radius ($R_J$) object
would yield a flux density ($\pi R^2 B_{\nu}d^{-2}$) of 14.4 and 0.8
$\mu$Jy at 4.5 $\mu$m for effective temperatures of 2000 and 750 K,
respectively, where $B_{\nu}$ is the Planck function. Such an object
is much fainter at the distance of P1121 (459 pc).  Therefore, the
excess emission and the flux modulation ($\sim$mJy at 4.5 $\mu$m) on
top of it most likely comes from dust emission and oscillations in its
thermal output in the system.

The low and relatively flat distribution in the 2012 disk light curve
around ID8 and the level of semi-regularity and complexity in 2013
successfully rule out many non-impact-related scenarios (for details,
see \citealt{meng14}).  The variations observed in the 2013 ID8 disk
emission required a large impact that produced an optically thick
cloud of glassy condensates and its subsequent orbital evolution. The
gradual flux decline in 2013 with a nominal timescale of one year is
consistent with a collisional cascade from parent bodies ranging from
a few times 100 $\mu$m to millimeter size \citep{meng14}. This size range of
condensates is consistent with the numerical model of spherule
formation in an impact-produced vapor plume \citep{johnson12}. This is
the main difference between the variable extreme debris disks and
typical debris disks where the collisional cascades start with at
least kilometer-size bodies whose collision timescales are long, resulting in
stable flux output for thousands to a few Myr.

\begin{figure} 
  \figurenum{7}
  \label{id8_orientation} 
 \epsscale{1.15} 
  \plotone{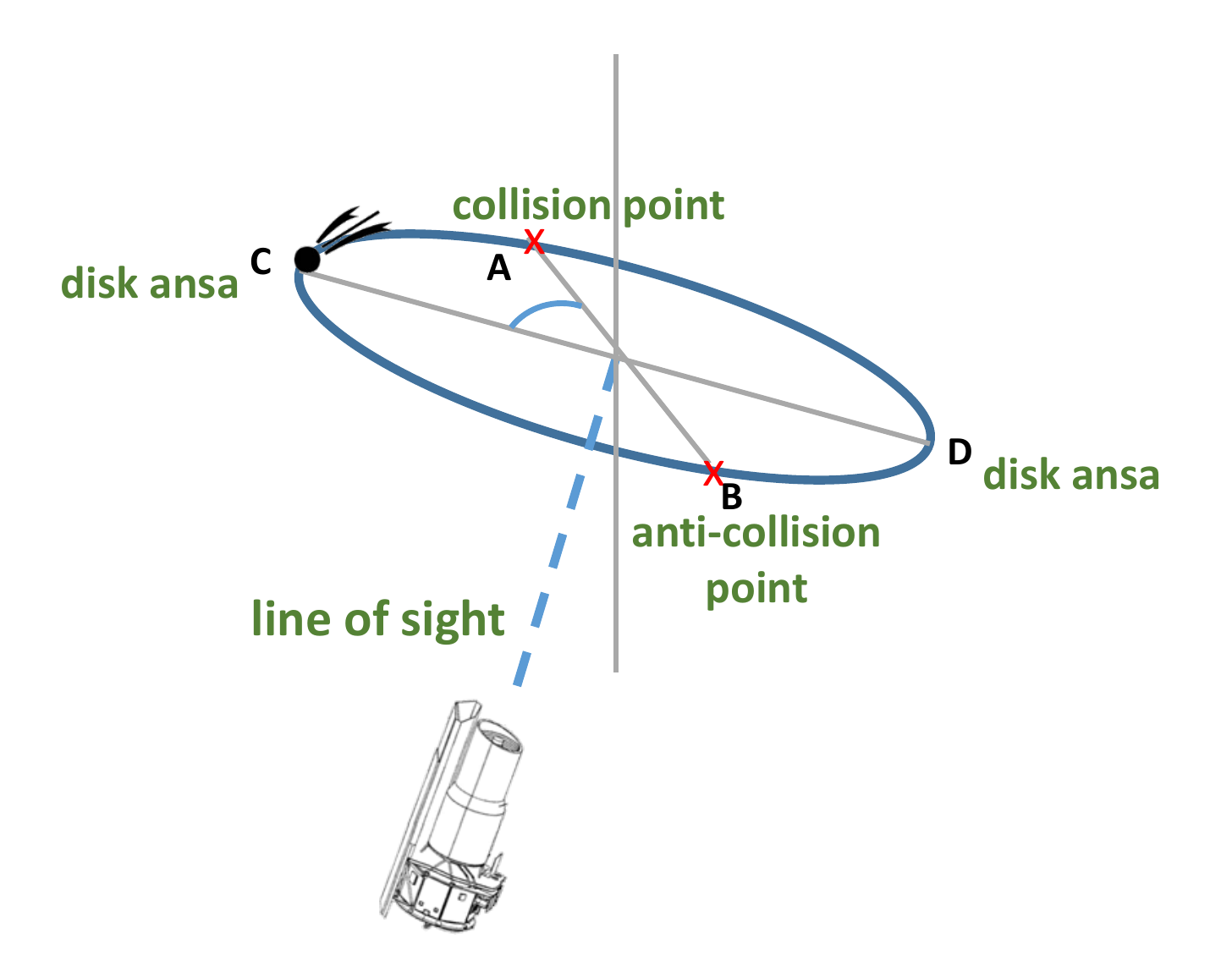}
  \caption{Sketch of the ID8 system to illustrate relative points
along the orbit of the impact-produced cloud, which is inclined from
the line of sight. The collision and anti-collision points are labeled
A and B, and C and D for the disk ansae. For an impact occurring 
halfway between the disk ansae, the angle (A to C) is
90\arcdeg. At these four spatially fixed places, the optical thickness
of the impact-produced cloud is expected to be highest (i.e., low flux
output in the light curve, details see \citealt{jackson19}).  }
\end{figure}

The flux modulations on top of the ID8 2013 gradual decline consist of
two intermixed periods (26$\pm$1 and 34$\pm$2 days, \citealt{meng14}),
which are too short for the orbital period of debris ($\sim$66--187
days at $\sim$0.32--0.64 au) inferred from SED modeling
\citep{olofsson12}. \citet{meng14} proposed that the modulations are
consistent with the changes of the projected area from an optically
thick cloud that is sheared along an eccentric orbit and is viewed
close to edge-on. The nearly edge-on geometry, which is consistent
with the inferred rotational axis from modulation of stellar spots,
naturally explains bi-periodicity because a cloud undergoing Keplerian
shear will be elongated in the orbital direction; therefore, at the
disk ansa (the end point along the disk major axis when viewed close
to edge-on, see Figure \ref{id8_orientation}) the cloud will be viewed
down its long axis, displaying its smallest sky-projected extent.  The
eccentric orbit and subsequent orbital evolution of the cloud result
in a complex periodicity with an actual orbital period of 75$\pm$5 days,
consistent with the SED-inferred debris location \citep{meng14}.

\citet{jackson14} provided a detailed description of
the dynamics of debris released by a giant impact. According to their
dynamical calculations, there are two spatially fixed locations for
the evolution of impact-produced debris: the collision point and the
anti-collision line (see their Figure 13).  The collision point is
where the impact occurred, which is a fixed point in space through
which the orbits of all of the fragments must pass because they
originated from there.  The orbital planes of the fragments thus share
a common line of intersection (line of nodes).  This leads to the
existence of the anti-collision line on the opposite side of the star
from the collision point along which the debris orbits cross
again. Detailed properties of the collision point and anti-collision
line and the evolution of their resultant asymmetric structures can be
found in \citet{jackson14}; here we refer to them as collision and
anti-collision points for simplicity.

Because the debris is funneled through a small volume at the collision
and anti-collision points, this naturally leads to a variation in cloud
cross section (i.e., brightness) with a period one-half of that of the
orbit for an optically thick cloud. These two effects (bi-periodicity
at disk ansae and bi-periodicity at collision and anti-collision
points) are independent of one another, with the relative phase
depending only on the orbital location at which the impact occurs (see
the illustration in Figure \ref{id8_orientation}).  For an edge-on
geometry, one would only expect a single periodic signal if the
collision occurred exactly halfway between the disk ansae, and the
true orbital period would be four times the single period. Similarly,
one would also expect a single period if the collision occurred at the
disk ansa, but the true orbital period would be twice the single
period instead. For a face-on geometry, there is no ansa effect, and so
there will only be a single bi-periodicity resulting from the
collision point/anti-collision points.  The combination of the disk
ansae and collision point/anti-collision points thus might naturally
explain the complex periodicity observed in the ID8 2013 light-curve
without invoking an eccentric orbit. The detailed evolution of the
light curve behavior, its dependency on geometry, impact condition, and
orbital eccentricity are further discussed in \citet{jackson19}.

\begin{figure} 
  \figurenum{8}
  \label{sed_dust_vs_grains} 
 \epsscale{1.15} 
  \plotone{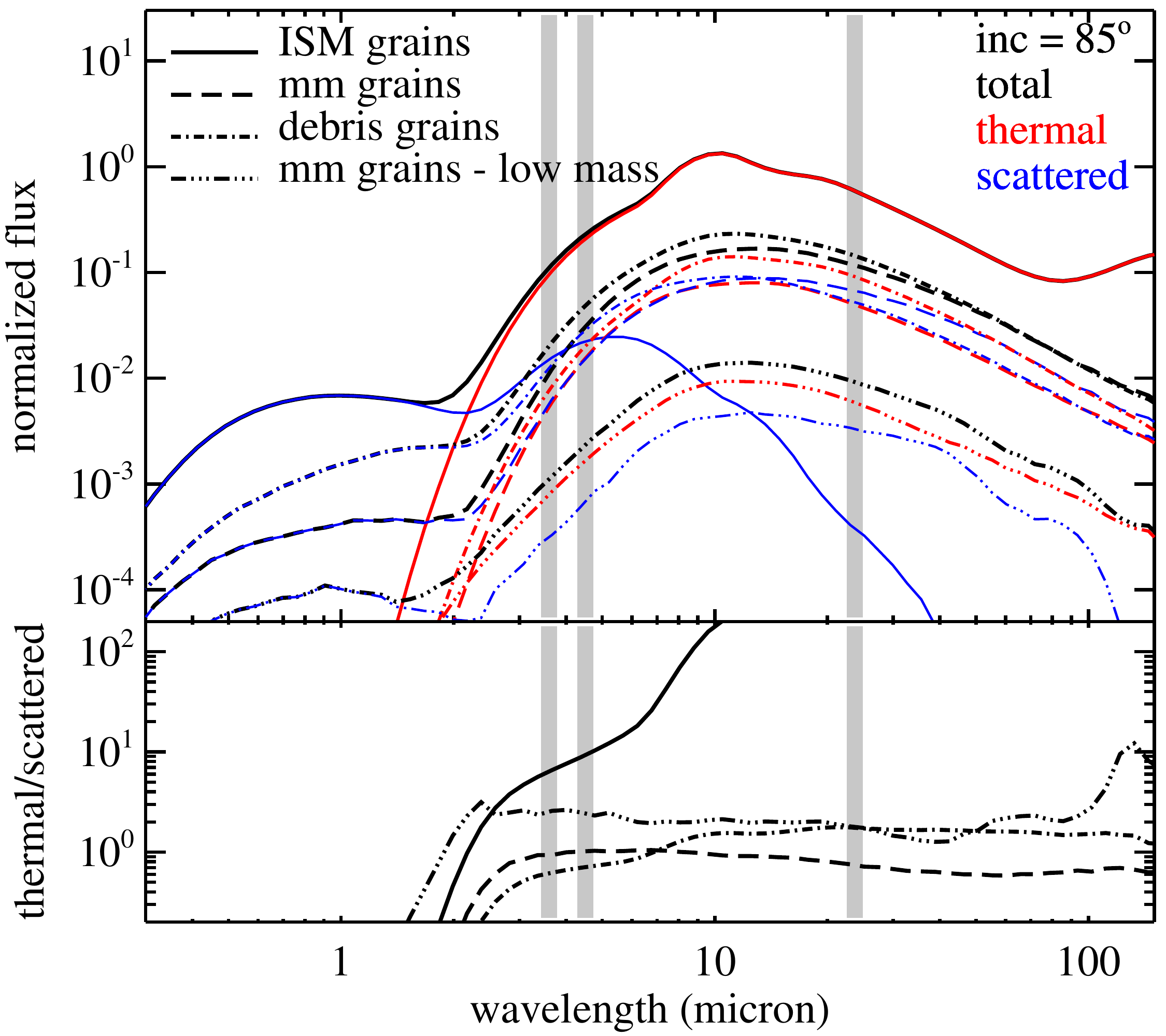}
  \caption{Example of the output SEDs for an impact-produced cloud
using three different grain size distributions: small ISM-like, large
mm size, and a wide range of sizes representing the typical size
distribution in nominal debris disks. The SEDs are normalized to show
the relative flux distribution, which is divided into two parts:
thermal and scattered emission, with the lower panel showing the ratio
of the two contributions. The density distribution and viewing
geometry of the debris is fixed in all cases of grain types, with a high dust mass
(very optically thick). One additional low dust mass model (by two
orders of magnitude) using the millimeter grains (dash-triple-dot line) is
also shown to illustrate the effect of optical depth (details see
Section \ref{3drt}). The vertical grey bars mark the wavelengths at
which variable disk emission is measured by existing observations. }
\end{figure}

\subsection{Radiation Transfer Calculations}
\label{3drt}

The geometric and dynamical model presented in the previous subsection
qualitatively describes the expected modulations from an optically
thick, impact-produced debris cloud. To translate such a model to the
actual measured flux, a full treatment of a radiative transfer model
is needed. In this subsection, we carry out 3D radiation transfer
calculations using the code developed by \citet{whitney13} that was adapted
by \citet{dong15} to perform protoplanetary dust disk simulations. The
radiative calculations include absorption, reemission and scattering
using the approximation of the Henyey-Greenstein function. The main
goal of these calculations is to demonstrate the feasibility of the
simple model and explore other parameters that might influence the
observed disk light curves.

\begin{figure*} 
  \figurenum{9}
  \label{sed_evo} 
 \epsscale{1.15} 
  \plotone{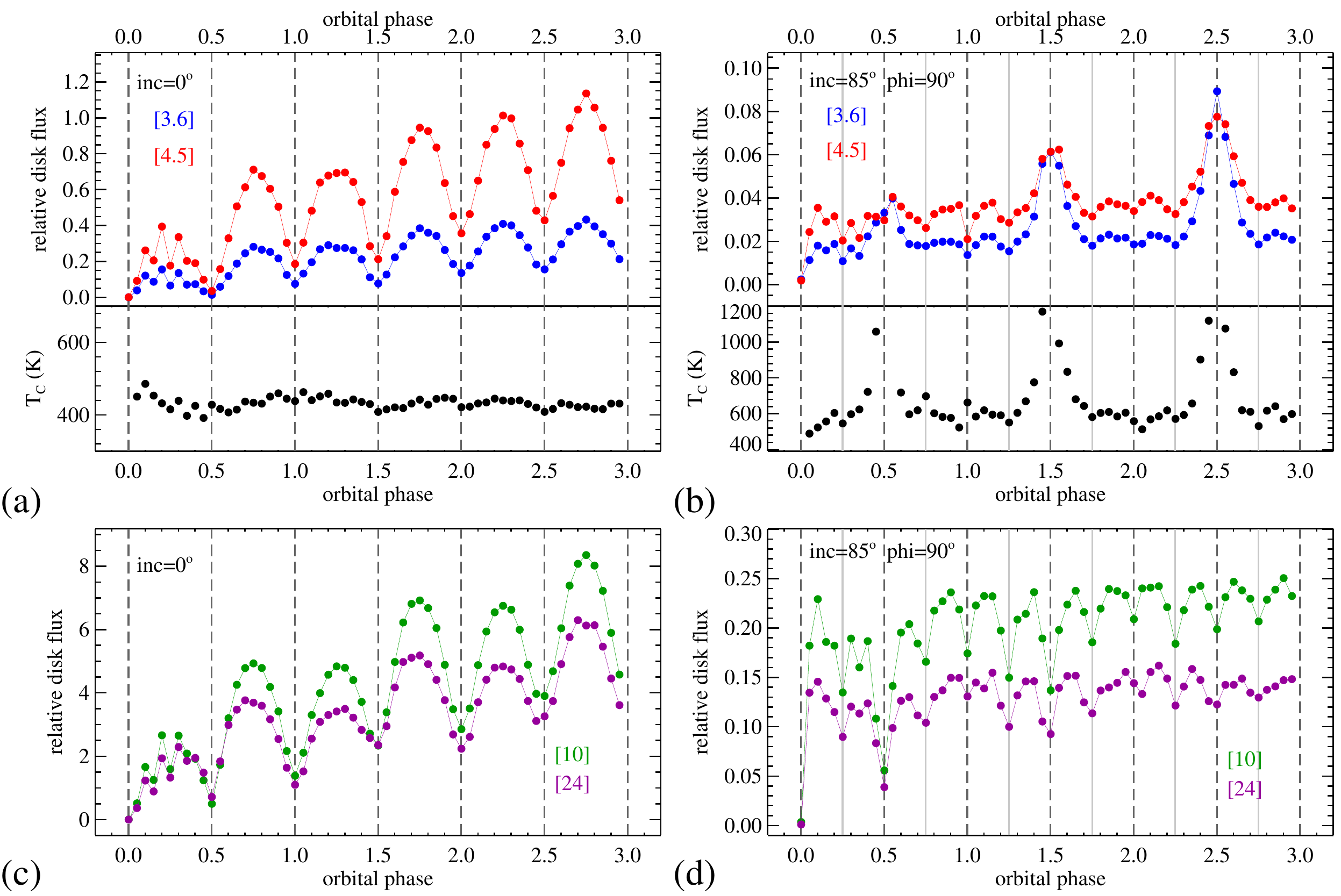}
  \caption{Simulated disk light curves of an optically thick cloud for
three orbits of evolution and at four selected wavelengths. Panels (a) and
(c) are for a high-mass cloud viewed face-on. As expected, the disk
emission is suppressed when the cloud passes the collision (phase of
integer numbers) and anti-collision (phase of half-integer numbers)
points, both marked by vertical dashed lines. Panels (b) and (d) are
similar plots, but viewed at 85\arcdeg\ from face-on and with 100 times
lower mass than in the face-on case. For this inclined case, the
collision point is set exactly between the disk ansae and behind the
star; therefore, the disk ansae are located exactly between the collision
and anti-collisional points, which are marked with vertical grey
lines. The color temperatures between [3.6] and [4.5] are also shown
in the bottom of panels (a) and (b). }

\end{figure*}

We first construct the 3D density distribution of the impact-produced
debris from the $N$-body simulations performed by \citet{jackson19} that
were qualitatively designed to fit the ID8 2013 disk
modulations. Details of the specific parameters in the numerical
simulation can be found in \citet{jackson19}. The 3D particle
distributions were recorded at 20 time steps per orbit with a total of
2.5$\times10^5$ particles. Each of the particles represents a
fixed fraction of the dust mass, depending on the assumed total
dust mass, as the input to the radiative transfer calculation. The
central heating source is assumed to be a main-sequence 5500 K star
with a stellar radius of 0.95 $R_{\sun}$. We first test the SED model
dependency on the chosen grain properties in terms of size range with
three different distributions: (1) interstellar median (ISM) grains: the size distribution
presented in \citet{kim94}  for the canonical diffuse interstellar
sightline (i.e, $R_V$ =3.1) as a representation for small grains, (2)
millimeter grains: 0.5--1 mm in a $-$3.5 power-law size distribution, and
(3) debris grains: 0.5 $\mu$m to 1 mm in a $-$3.5 power-law size
distribution. All three size distributions have the same mixture of
compositions as described by \citet{kim94}, containing silicate,
graphite, and amorphous carbon (see Section 2.2 in \citealt{dong12}
for more details). To test the feasibility of the simple geometric
model, we assume that each particle represents the same grain sizes at all
times, i.e., no collisional evolution within the cloud. An
example of the resultant SEDs is shown in Figure
\ref{sed_dust_vs_grains}, where the total dust mass of the cloud is set
to be 2.5$\times10^{-4}$ M$_{\oplus}$ (i.e., very optically thick),
viewed at an inclination angle of 85\arcdeg\ from face-on, and after
one orbital evolution since the impact. To test the optical depth
effects, we also compute the SED of millimeter-size grains with a mass two
orders of magnitude lower than the previous value. The output SED is
divided into two parts: thermal component and scattering component,
including both scattered starlight and the cloud emission. With the
fixed viewing geometry, the relative contribution of these two parts
depends sensitively on the grain properties and optical depth, as shown
in the bottom panel of Figure \ref{sed_dust_vs_grains}. Except for the
very small ISM grains, the scattered component is not negligible and
dominates for wavelengths shorter than $\sim$2 $\mu$m in the final SED
output.

Figure \ref{sed_evo} shows the flux evolution of the cloud over three
full orbits at four wavelengths of interest: [3.6], [4.5], [10], and
[24]. All other parameters of the cloud are fixed (using the millimeter 
grains), except for the viewing angles: face-on (an inclination of
0\arcdeg) and close to edge-on (an inclination angle of 85\arcdeg),
and the total dust mass: 2.5$\times10^{-4}$ M$_{\oplus}$ (high mass) for the
face-on case and 2.5$\times10^{-6}$ M$_{\oplus}$ (low mass) for the inclined
case. The initial point of the orbital phase is defined at the
collision point (phase of 0.0). The orbital phase of 1.0 is at the
same point but after one orbit of evolution, and the orbital phase of
1.5 is its corresponding anti-collision point. For the face-on case,
the cloud is so optically thick that the resultant SEDs are very close
to the projected, geometric cross section of the cloud at different
orbital phases -- the flux is lower at the collision and
anti-collision points than at their prior adjacent phases.  There
is a gradual rising trend in the mid-infrared flux due to Keplerian
shear that increases the surface area of the cloud over time.  For the
inclined case, the collision point is set exactly between the disk
ansae behind the star, i.e., the disk ansae are at the orbital phases
of 0.25 and 0.75 after the impact. The evolution of the inclined disk
SEDs is more complex than that of the face-on case. At the wavelengths
where the scattered component is important, the disk flux swings
greatly and reaches maximum at the anti-collision point because the
grains used in the radiative calculation are strongly forward-scattering. 
Such a large flux swing is not seen for the face-on case
at similar wavelengths because the cloud has the same scattering
angle. This explains that the flux of the cloud in the inclined case
reaches local maximum instead of minimum at the anti-collision point
(phases of 0.5, 1.5 and 2.5 in Figure \ref{sed_evo}) at 3.6 and 4.5
$\mu$m where the scattering component from the starlight is
important. This is also consistent with the jumps in the observed
color temperatures (the bottom panel of Figure \ref{sed_evo}b) for the
inclined geometry. At longer wavelengths at which the scattered starlight 
is not as important, the flux of the cloud drops whenever it
passes the collision and anti-collision points and disk ansae (Figure
\ref{sed_evo}d).  In all the radiation transfer calculations, the
cloud is placed at the same radial location from the star. When we exclude the
large color temperature swings due to scattering, the derived color
temperatures between [3.6] and [4.5] differ by no more than 100 K
between the high- and low-mass clouds, suggesting that the
[3.6]--[4.5] color is not sensitive to cloud location under an
optically thick condition.

As mentioned in Section \ref{debrislocation} (see Figure
\ref{id8_p1121sed}), the dust traced by the warm {\it Spitzer} data
might be separate from the dust that emits the prominent solid-state
features (i.e., the latter may arise from a reservoir of
planetesimals); therefore, it is not surprising that the computed SEDs
(Figure \ref{sed_dust_vs_grains}) do not resemble those in Figure
\ref{id8_p1121sed}. The fact that there is relatively little change in
the observed color temperatures might indicate that the grains are not
as forward-scattering as the model grains. We also note that the optical
depth and the collisional evolution within the cloud might also affect
the observed color temperatures. In summary, our pilot study with the
full radiation transfer calculations qualitatively confirms the
expected modulations at the disk ansae and collision and
anti-collision points.  To better extract more information about the
system, such as the required minimum dust mass to produce a modulation
and a better constraint on the system's inclination angle, a full
exploration of other parameters to match the observations
quantitatively will be presented in a future work.

\subsection{Application to the modulations in the ID8 disk light curves }
\label{modulations}

\begin{figure*} 
  \figurenum{10}
  \label{diskphase20134} 
  \epsscale{1.15} 
  \plottwo{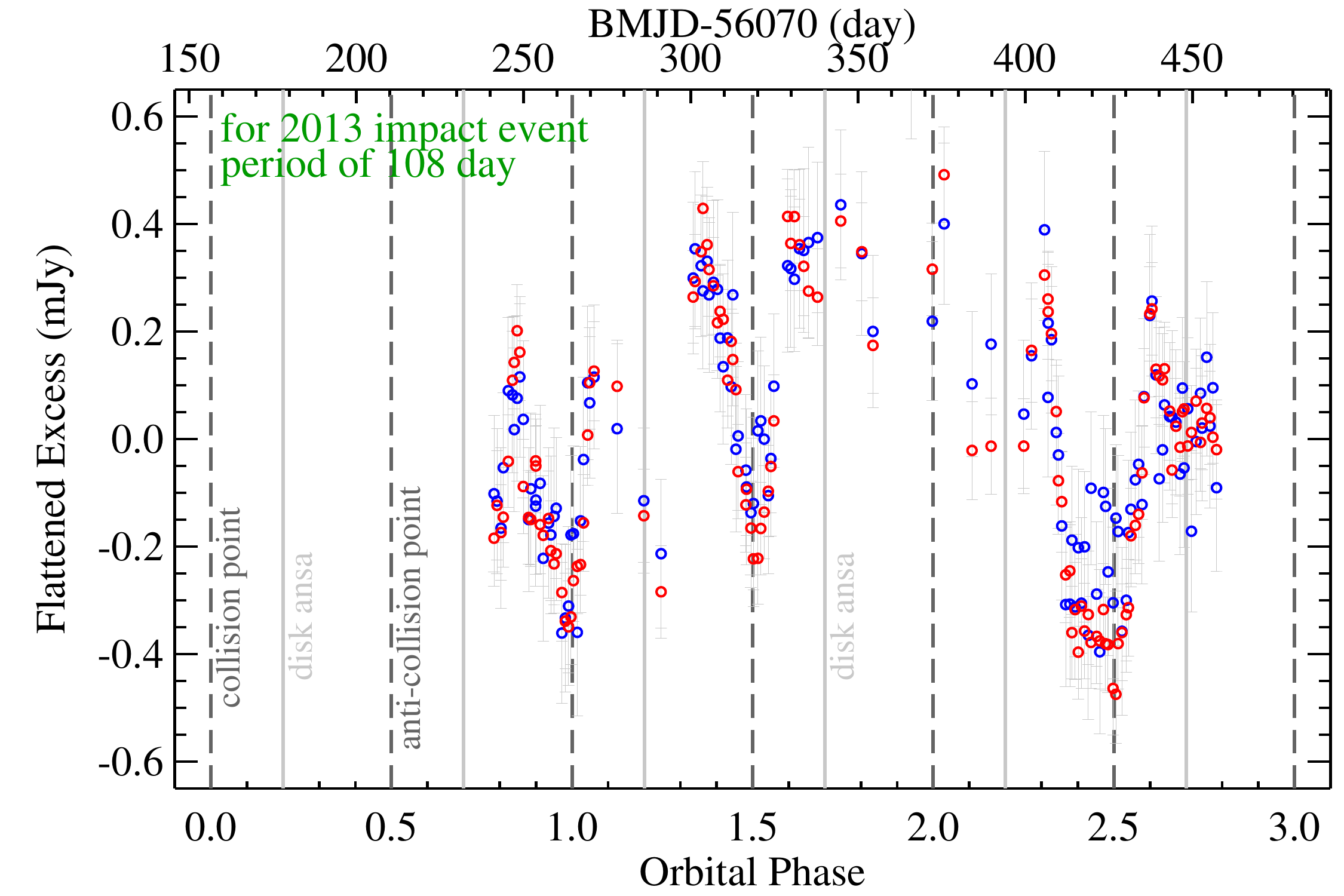}{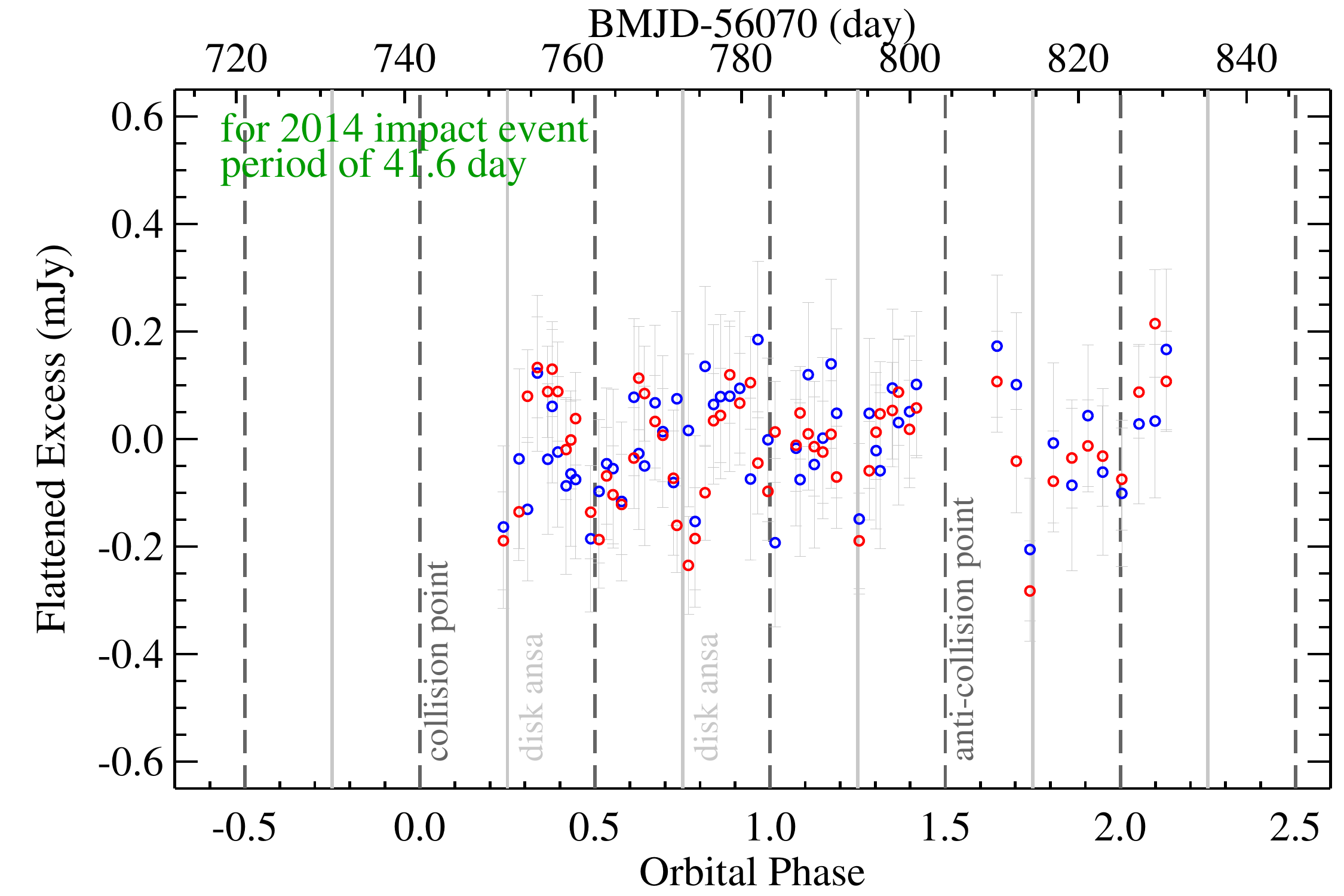}
  \caption{Left (right) panel shows the ID8 2013 (2014) flattened light
curve expressed in the orbital phase by dividing the true orbital
period. The true orbital period is 108 days for the 2013 modulation,
but 41.6 days for the 2014 modulation. In both panels, the vertical
dashed lines mark the phases for the collision and anti-collision points, while the
vertical grey lines mark the phases for the disk ansae
(details see Section \ref{modulations}).}
\end{figure*}

One of the conclusions from the previous subsection is that the lowest
flux always occurs at the collision and anti-collision points, with
the second lowest flux at the disk ansae right after an impact in an
inclined geometry. Encouraged by these results, we first tried to
identify the collision and anti-collision points and determined the
half orbital period between them in the ID8 2013 flattened light curve
using the phase dispersion minimization approach
\citep{stellingwerf78}. When we assume that the first large dip (flux minimum)
observed on d.d.\ 264 is due to one of the collision points, the next
large dips are likely associated with anti-collision and collision
points. The true orbital period of the cloud should always be twice 
the half orbital period between the collision and anti-collision
points. Looking at the dips in 2013, we determined that the half orbital
period between collision and anti-collision points is 54 days,
suggesting that the true orbital period is 108 days.  The left panel of
Figure \ref{diskphase20134} shows the 2013 phased flattened light
curve where the easily identified large dips are all lined up with the
collision and anti-collision points, suggesting the robustness of this
period. An orbital period of 108 days indicates that the impact
occurred at a distance of 0.43 au from the star, within the expected
debris location.  Identifying the dips due to disk ansae is trickier,
especially if the particles within the cloud are collisionally
evolving (i.e., the distribution of the particle sizes is
evolving). By examining the nearby, secondary dips around the
identified collision and anti-collision points, we then identified
that the disk ansae are likely at the orbital phases of 0.2 and 0.7
(using the dips near d.d.\ 286 and 449). These orbital phases mean
that the cloud reached the disk ansae 21.6 days after passing the
collision and anti-collision points (from A to C or B to D in Figure
\ref{id8_orientation}) . After this, the cloud reached the next
following collision and anti-collision points (from C to B or D to A
in Figure \ref{id8_orientation}) after 54 $-$ 21.6 $\sim$33
days. Ideally, in a well-sampled light curve, we would expect to find
possible dominant periodic signals of 108, 54, 21, 33, and 75 days. As
shown in Figure \ref{diskphase20134}, our sampling is relatively poor
when the cloud reached the disk ansae. The allowable margins for the
phases of the disk ansae are large (not as robust as the collision
and anti-collision points).  The sampling effect combined with the
possible collision evolution within the cloud results in a situation
that the periods of 26$\pm$1 and 34$\pm$2 days are dominant in the
periodogram analysis, as reported by \citet{meng14}.

The phase difference between disk ansa and collision point
suggests that the angle between the collision point and the disk ansa
is about $\sim$70\arcdeg.  In principle, the first large dip observed in
a light curve could be the collision or anti-collision point after an
integer number of orbits after the impact. If the first large dip on 
d.d.\ 264 was associated with the collision point, it must be
associated with the phase of 1.0 (exactly one orbital evolution)
because the impact event had to occur during the {\it Spitzer}
visibility gap between d.d.\ 85 and 240, i.e., on BMJD 56227
(2012 October 26, or d.d. 157) given the orbital period of 108
days. Alternatively, the first large dip could also be associated with
the anti-collision point, i.e., orbital phase of 0.5 (any integer
number of 0.5 would place the impact event during the 2012 flat light
curve) with the impact event on d.d.\ 210. However, this seems slightly 
unlikely because the warm {\it Spitzer} observations trace small
grains, and it takes time to produce them in an impact-produced cloud
through collisional cascades (details see Section
\ref{codeMresults}). In summary, the short-term modulation observed in
the 2013 data is consistent with an impact event that occurred in late
2012 (called the 2012 impact event) at 0.43 au from the star.

The modulation period in 2014 is not only very different from the period
in 2013, their associated long-term trends are also in stark contrast:
downward vs.\ upward. This argues for a different origin from the 2012
impact event.  The single modulation period in the 2014 data suggests
that the disk ansae are exactly at the halfway point between the
collision and anti-collision points (i.e., A to C, C to B, B to D, and
D to A in Figure \ref{id8_orientation} are all 10.4 days), and the
angle between the collisional point and disk ansae is 90\arcdeg.  The
true orbital period is then 41.6 (4$\times$10.4) days for the 2014
impact event, implying an orbital distance of 0.24 au. The short-term
modulations in 2013 and 2014 are caused by two different optically
thick clouds produced by two distinct impact events.  To further test
this hypothesis, we phased the 2013 and 2014 light curves together
with the same period of 108 days and impact date, and found no
corresponding dips with the expected collision and anti-collision
points in the 2014 light curve. This corroborates that there were 
two independent impact events.

For the 2014 impact event, we tentatively set the impact to occur on
d.d.\ 742, therefore the dip on d.d.\ 763 represents the
anti-collision point. The right panel of Figure \ref{diskphase20134}
shows the 2014 phased flattened light curve. The real date for this
2014 event is likely to be earlier given the low {\it WISE} flux on
d.d.\ 725 (Figure \ref{timeseries_id8_p1121}a). We compared the
2015--2017 flattened light curves in phase space with the light curve of 2014 
using the same period of 41.6 days to identify additional modulations
that might be produced by the same cloud. The flux dips on d.d.\ 1012,
1033, and 1073 (all obtained in early 2015) are likely associated with
the orbital phases of 6.5, 7.0, and 8.0 (marked in Figure
\ref{id8_2013567}b) from the 2014 event. The flux variation becomes
more stochastic afterward, except that a deep dip on d.d.\ 1438 (in
2016) might be associated with the orbital phase of 16.75 due to one
of the disk ansae. The overall short-term temporal behavior is
consistent with the expected evolution -- the impact-produced clump in
the disk lasts for $\sim$10 orbits when the disk flux modulation is
strong and observable during this clump phase
\citep{jackson14,jackson19}.

\subsection{Application to the modulation in the P1121 disk light curves }
\label{modulations_p1121}

For P1121, a modulation with a single period of 16.7 days is seen in
the 5 yr of {\it Spitzer} data. From the ground-based optical
monitoring (presented in Section \ref{optical}), no periodicity due to
rotating stellar spots is found; i.e., the orbital plane of the debris
is not likely close to edge-on, as in the case for ID8. From the SED
models presented in Section \ref{debrislocation}, the debris location
is estimated to be at $\sim$0.2--1 au (orbital periods of
$\sim$33--365 days). If the disk light-curve modulation in P1121 is
caused by the orbital evolution of the impact-produced cloud, the true
orbital period is likely to be 2$\times$16.7 = 33.4 days (an impact at
0.2 au) for a face-on geometry, or 4$\times$16.7 = 66.8 days (an
impact at 0.32 au with the collision point halfway between the disk
ansae) for a close to edge-on geometry. Both are within the estimated
debris location. However, it is difficult to have the cloud remain in
the clump phase for more than 30--60 orbits after the impact,
especially after the excess emission reaches the background flux level
(i.e., the impact-produced cloud has dissipated and/or merged with the
existing debris belt). We note that the amplitude of the modulation
($\pm$0.08 mJy at 4.5 $\mu$m) observed in P1121 is very similar to the
amplitude created by the ID8 2014 impact event, which only lasted for
$\sim$10 orbital periods (less noticeable in late 2015). The longevity
of the short-term modulation in P1121 argues against the post-impact
possibility. We discuss other possible scenarios for the
short-term modulation in Section \ref{otherscenarios}.

\section{Interpretation: Long-term Variability}
\label{interpretation_long-term}

The debris generated by a violent impact is characterized by two
different populations: (1) the ``vapor population'': the
escaping fragments produced from the recondensation of melt or vapor;
and (2) the ``boulder population'': the fragments escaping in
the unaltered solid state.  The ratio between the two populations
depends on the impact conditions, i.e., the vapor/melt population can be
the dominant product in a hypervelocity impact (e.g.,
\citealt{benz07}; \citealt{svetsov16}; \citealt{lock18}).  Both
populations, once in a circumstellar orbit, will start to collide and
produce new, smaller fragments that grind down to small dust that
emits efficiently in the infrared. The geometric and dynamical model
presented in Section \ref{interpretation_short-term} does not
differentiate these two populations as long as the cloud is optically
thick to produce the short-term light curve modulation. That is, it
assumes an appropriate configuration for the cloud without considering
collisional evolution within the cloud. The extended time coverage
reported in this paper documents the long-term (yearly) evolution of
the infrared output from these two systems. To interpret the long-term
evolution of an impact-produced cloud, we do need to consider the
collisional evolution within the cloud -- how small grains that are
probed by the infrared observations are generated (referred to as the 
``buildup'' phase) and the associated mass depletion resulting in the
infrared flux decay from the system (referred to as the ``decay'' phase).

\begin{figure*} 
  \figurenum{11}
  \label{crossection} 
  \epsscale{1.15} 
  \plotone{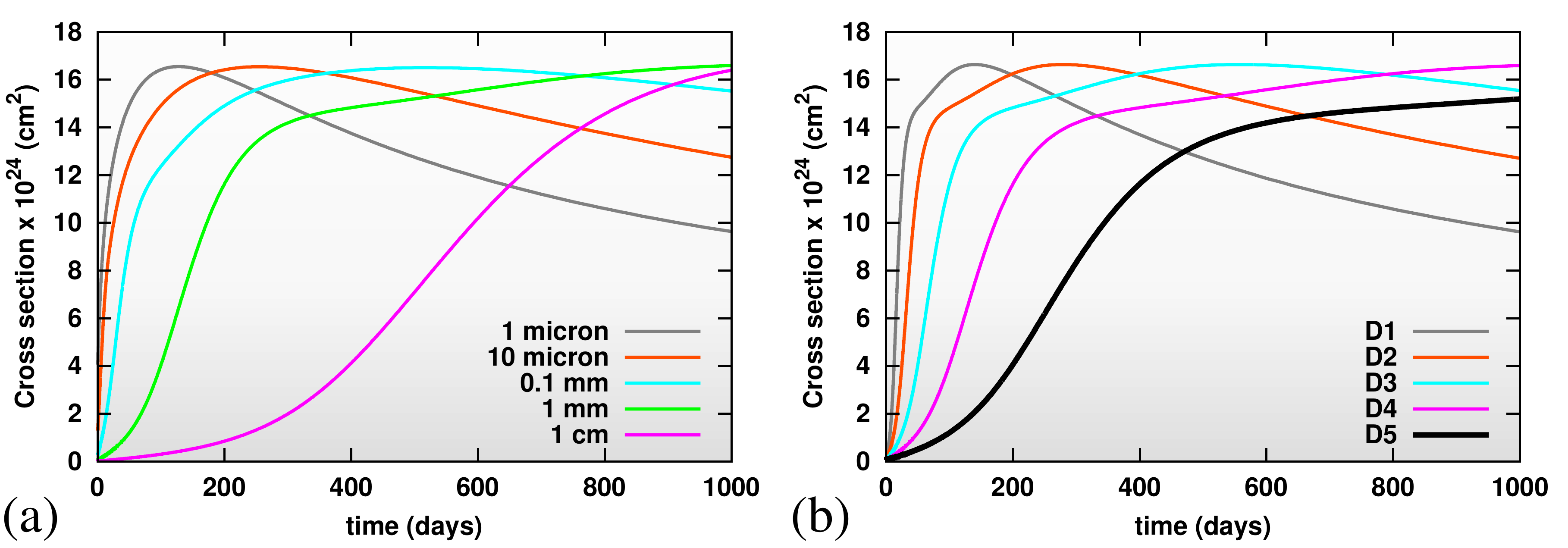}
  \caption{Evolution of the total cross section for a swarm of
particles at 0.24 au from a 0.9 $M_{\sun}$ star with a total mass of
0.28 M$_{Moon}$ and $s_{max}$=100 km. The left panel (a) shows the
buildup rates with various initial minimum cutoffs ($s_{min}$ listed
in the plot) in a fixed volume, while the right panel (b) shows the
buildup rates with five initial densities: D1 to D5 = [144, 72, 36,
18, 9]$\times10^{-10}$ kg~m$^{-3}$ in the swarm with fixed $s_{min}=
1$ mm and $s_{max}=100$ km and total mass by varying volume of
collisional space.}
\end{figure*}

The speed of the collisional evolution and therefore the flux changes
are determined by the collisional timescales in the populations of
different-sized particles in the cloud. A population of small
particles reaches collisional equilibrium much faster than a population of
large particles; as time goes by, larger sized populations that are
longer lived will feed these cascades over time. Therefore, the collisional
evolution of the recondensed, boulder, or a mixture of the two
populations would be very different. Given the limited information we
have for the condition of an impact-produced cloud, a full exploration
of the possible parameters involving dynamical (previous section) and
collisional evolution is beyond the scope of this paper.  In Section
\ref{codeMresults} we use the collisional cascade code developed by
\citet{gaspar12} to qualitatively illustrate the collisional evolution
of a swarm of particles in a particle-in-a-box (1D) approach. We note
that the large density enhancement at the collision point (as
described in Section \ref{basicideas} and \citealt{jackson19}) means
that a direct translation is difficult from a 1D or any analytical
approach to a realistic confined region (an impact-produced cloud). We
only qualitatively interpret the observed long-term flux behavior for
both systems in Section \ref{int_long-term} to assess the impact
scenario.

\subsection{Collisional Evolution of an Impact-produced Cloud}
\label{codeMresults}

The numerical code by \citet{gaspar12} is designed to model
collisional cascades in debris disks describing both erosive and
catastrophic collisions among particles statistically in a limited
volume and treating the orbital dynamics of the particles in an
approximate fashion. We use this code to model collisional cascades to
explore the expected dust cross-section (i.e., flux) evolution using
various initial conditions (initial density and size distribution of
the fragments).  All simulations were run around a 0.9 $M_{\sun}$ star
with a swarm of particles at 0.24 au. The initial volume of the swarm
is set to 0.0037 au$^3$ (i.e., $\Delta r / r \sim 0.01$ for $r$ = 0.24
au). At time zero, a swarm of particles has initial sizes ranging
from minimum radius ($s_{min}$) to maximum radius ($s_{max}$) in a
size distribution slope of $-3.65$ (an adopted value to roughly match
the steep\footnote{We note that the small difference in the power
index in such a steep size distribution does not affect our
qualitative conclusion.} size distribution from hypervelocity impact
experiments \citealt{takasawa11}). We set the collisional
velocities\footnote{See \citet{mustill09} and \citet{gaspar12} for the
definition of collisional velocities in a swarm of particles.} to 3 km
s$^{-1}$, which is 5\% of the Keplerian orbital velocity at 0.24
au. This value is slightly lower than the 10\% value assumed for
typical debris disks. These particles likely originate from a single
body and therefore are on similar orbits, requiring a reduced
collisional velocity.

We first explore the buildup phase using a swarm of particles with a
fixed 100 km for $s_{max}$ and various $s_{min}$ cutoffs.  Figure
\ref{crossection} shows the time evolution of the collisional system,
expressed as the total cross section ($\sigma_T$) integrated over all
sizes from the blowout size (set to 0.5 $\mu$m) to the maximum
size. Because the emitting flux is closely proportional to $\sigma_T$,
the flux evolution of an impact-produced cloud qualitatively follows
the evolution of the total cross section.  As shown in Figure
\ref{crossection}a, the smaller the initial minimum size in a swarm,
the faster the cloud reaches the maximum in the total cross section
(i.e., the quasi-static state collisional cascades). The rate of
generating small grains (i.e., increasing the total cross section,
called the buildup rate) depends sensitively on the minimum size of
the fragments. A slow buildup rate over a course of $\sim$2 yr
(similar to the rise between the end of 2013 to early 2015 in the ID8
system) suggests a minimum size between 1 mm and 1 cm. However, it is
difficult to determine the exact $s_{min}$ because the buildup rate
also depends on the initial cloud density. Figure \ref{crossection}b
shows the evolution for a swarm of particles with a fixed size
distribution ($s_{min}=1$ mm and $s_{max}=$ 100 km), but at different
initial densities (by changing the total volume and keeping the same
total mass). As expected, the higher the initial density, the
faster the buildup rate, i.e., a faster time to reach the
maximum part of the curves. We found that as long as the lower size
of the fragments exceeds the centimeter size, the buildup rates are
not as sensitive to the lower size limit itself as they are to the
initial cloud density.  The initial density and the minimum size of
the fragments are degenerate in determining the buildup rate; however,
a lack of micron-sized fragments in the initial size distribution
is necessary to reproduce the slow (multiyear) buildup in a swarm of
colliding bodies.

A violent impact is likely to produce various combinations of the
vapor and boulder populations (e.g., \citealt{svetsov16}),
resulting in different size distributions for the initial
fragments. To qualitatively explore the possible collisional outcomes
for a swarm of impact-produced fragments, we ran a series of
simulations by mixing different populations, as described below. In
these simulations, the vapor population is defined as fragments with
sizes of 1--5 mm, while the boulder population is defined as fragments
larger than 5 mm with various maximum size cutoffs ranging from 10 m
to 1000 km, which determines the total mass (i.e., the mass of the
swarm is proportional to $s_{max}^{0.65}$ for a size slope of
$-$3.65). The exact division between vapor and boulders has little
impact on the collisional calculation because they are all treated as
particles with size-dependent strengths. For a mixture of the
populations, we mean that the size distribution in the region of small
sizes has a jump (not a continuous power-law distribution),
representing the additional vapor population. The initial volume and
collisional velocities in the swarm are fixed as stated before, i.e,
the density of the cloud is not fixed.

\begin{figure*} 
  \figurenum{12}
  \label{codeMsimulations} 
  \epsscale{1.15} 
  \plotone{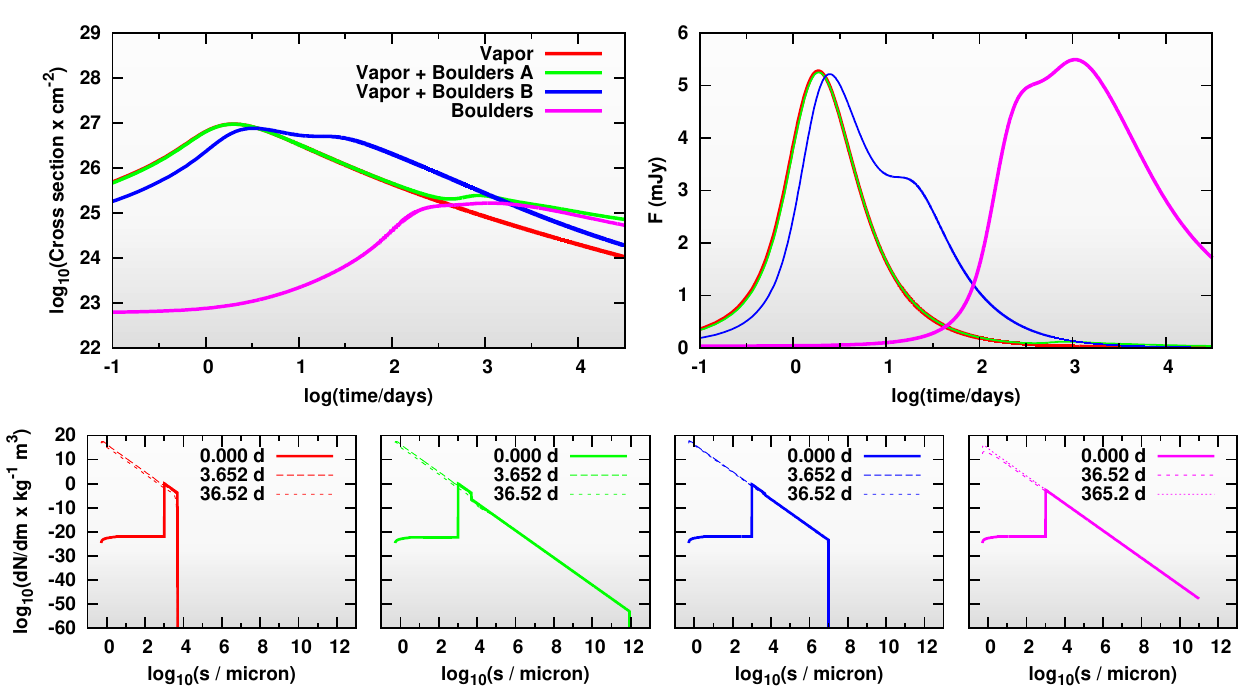}
  \caption{Expected collisional evolution around a 0.9 M$_{\odot}$
star for a swarm of particles within a fixed volume (0.0037 au$^3$ at
0.24 au) with characteristic collisional velocities of 3
km~s$^{-1}$. The upper left panel shows the evolution of total dust
cross section integrated from $s_{min}$ to $s_{max}$, while the upper right
panel shows the expected 4.5 $\mu$m flux assuming an optically thin
condition with an instantaneous loss of grains smaller than the
blowout size. The bottom panels show the size distribution of the
swarm initially (solid lines) and at two selected days (long-dashed
and dashed lines). Different colors represent the four different
initial conditions: total mass and particle size distribution (details
see Section \ref{codeMresults}).  }
\end{figure*}

We test four different initial
conditions: (1) vapor only, (2) vapor plus boulders up to 1000 km in
radius (called vapor+boulder A), (3) vapor plus boulders up to 10 m in
radius (called vapor+boulder B), and (4) boulder only (up to 100 km).
For the mixture cases, 20\% and 10\% of the total masses are in the
vapor form for vapor+boulder cases A and B, respectively.  Figure
\ref{codeMsimulations} shows the results of these simulations where
the collisional evolution of the swarm is shown not only as the total
cross section, but also as the size distribution at three selected
days after the impact. Furthermore, the expected 4.5 $\mu$m flux using
the optically thin assumption is also shown as one of the panels in
Figure \ref{codeMsimulations}. Although the optically thin flux
calculation is not truly representative of the actual observed flux
(especially in the early evolution due to the optical thickness), it
does reflect the expected flux drop once the system reaches
quasi-static collisional equilibrium when the largest fragments start
to participate in collisional cascades. In the optically thin flux
calculation, we also take into account the grain-size-dependent
absorption coefficients and their resultant thermal-equilibrium
temperatures by adopting the composition of astronomical silicates. We
further adjust the initial total mass in the swarm so that the peak
4.5 $\mu$m flux reaches $\sim$5 mJy during the evolution.

For the vapor-only model, the total mass is 0.176 M$_{Moon}$, i.e., a
very high particle density in the initial volume. As a result, the
initial buildup phase is short; it reaches the quasi-steady state
collisional equilibrium within a few days and follows a fast drop-off
afterward (a rise-and-fall behavior). We note that the drop-off rate
in the optically thin flux is artificially enhanced simply because the
collisional code does not track the blowout grains in any timely
fashion (the blowout grains are assumed to be instantly lost from the
system, while they should be moving outward with a terminal velocity
depending on their sizes). 

For the vapor+boulder model A, the total
mass of the system is 0.85 M$_{Moon}$ and 20\% of the mass is in 
vapor form (similar to the vapor-only model). Its initial evolution
within a few hundred days is the same as the vapor only model, but a
secondary buildup occurs near 500 days and reaches the maximum near
1000 days, followed by a very slow decay afterward. In this
vapor+boulder A model, the first fast buildup is totally dominated by
the vapor population where the total cross section has a quick
rise-and-fall behavior just like the vapor only model, while the
second rise comes from the slow buildup of the boulder population. A
similar behavior is also seen in the vapor+boulder B model where the
total mass of the system is 0.76 M$_{Moon}$ and 10 \% of it is in the
vapor form.  Because the mass of the vapor is roughly half of that of the
vapor-only model (i.e., slightly lower density in the same volume), it
takes slightly longer for the initial buildup than that of the
vapor-only model. Although the total mass in the boulder population
between the two vapor+boulder models is similar ($\sim$0.68
M$_{Moon}$), the particle density is different because of the different
maximum sizes (1000 km in vapor+boulder A, but 10 m in vapor+boulder
B), i.e., the particle density is much higher when the largest
fragment is small. As a result, the secondary buildup due to the
boulder population is also much faster, it occurs within $\sim$20
days. The total cross section then follows a faster drop-off than
the one from the vapor+boulder A model (the slope of the blue curve
is steeper than the slope of the green curve after 1000 days in the
upper left panel of Figure \ref{codeMsimulations}). This is because
the drop-off rate is mostly governed by the collisional timescale of
the largest fragments, i.e., the larger the size, the slower the
drop-off rate\footnote{We note that the drop-off rate also depends on
the initial density. In an extremely high-density environment (such as
the D1 curve in Figure \ref{crossection}b), the decay rate could be as
short as $\sim$10 yr for a swarm of particles with $s_{max}$= 100
km.}.  

For the boulder-only model, the total mass is 0.28 M$_{Moon}$; 
0.1\% of the grains have sizes in the range of 1--5 mm (consistent with
no vapor population). The buildup phase takes much longer (a few 100
to some 1000 days) due to the low initial density. The drop-off rate is
similar to that of the other boulder models, dominated by the collisional time
scale of the largest fragments (100 km in this boulder-only model).
We note that a similar buildup and drop-off curve can be obtained if
one reduces the largest fragments to 10 km in the boulder only model
(instead of 100 km) and the same particle density is maintained by increasing
the volume. In this case, the total mass is 0.13 M$_{Moon}$, which is roughly
half of that of the original boulder-only model. Hence, the total mass of the
swarm cannot be well constrained using these 1D simulations.

We can draw some basic conclusions from these simulations. In a
high-density environment (fragments produced in a violent impact and
in a cloud that is likely to be initially optically thick to the starlight), 
the collisional evolution of vapor condensates is fast,
which quickly generates many small grains and reaches quasi-static
state colllisional equilibrium within a few days.  In contrart,
the evolution of a boulder-only population is rather slow, and could
take up to months to generate enough small grains, resulting in an initially slow
rise. Once the collisional system reaches a quasi-static
state and starts the mass depletion in the largest fragments, the
total cross section starts to fall with the rate depending on the
collisional timescale of the largest fragments. The flux decay
timescale is much faster for the vapor condensates than that of large
boulders. For an impact that produces mostly vapor condensates, a fast
rise-and-fall behavior is likely to appear in the observed flux. For
an impact that produces large boulders, the initial flux rise could be
slow, depending on the minimum size of the boulders and initial
density, followed by a flux decay that is also much slower than that
of vapor condensates. The evolution for a high-density cloud that has
a mixture of vapor and boulder populations is likely to be initially driven by
the vapor population initially (a fast buildup), and a possible
secondary buildup from the boulder population with the timescale
depending on the initial density of the boulder population (the higher the
density, the faster the secondary rise).  We stress that the
collisional evolution is sensitive to the initial density (i.e., the
volume). Therefore, the timescales of different evolution models
(buildup and drop-off) should be taken qualitatively (not
literally). We also note that our simulations have many fixed
parameters (initial location, volume, and collisional velocities) that
are not fully explored. A full exploration of the required parameters
with a proper radiative transfer calculation to match the long-term
flux evolution will be presented in a future study.

\subsection{Matching the Long-term Behaviors in ID8 and P1121}
 \label{int_long-term}

The previous subsection qualitatively demonstrated the expected
long-term behaviors due to the collisional evolution within an
impact-produced cloud. In this subsection, we will discuss the observed
long-term flux trends in both systems to assess
whether the trends are consistent with the post-impact nature.

\subsubsection{Single Large Impact in the P1121 System Prior to 2012}

We start with the P1121 system because its disk light curve behavior is
less complex -- a flux decay since 2012 with a
timescale of $\sim$one year and reaching a background level ($\sim$1.2
mJy at 4.5 $\mu$m) that was reached since 2015 (Figure 4). The long-term disk variation
in the P1121 system is consistent with a hypervelocity impact that
occurred prior to 2012. As demonstrated in the simulations presented
in Section \ref{codeMresults}, the year-long flux decay is consistent
with the rapid collisional evolution from a swarm of particles that condensed
from vapor. The mid-infrared spectrum taken in 2007 provides some limits on
the location (0.2--1.6 au) and the amount of dust
($\sim$9$\times10^{-6} M_{\oplus}$) in the P1121 system (see Section
3.3). When we assume that the short-term modulation (16.7 days) on top of the
flux decay is due to the orbital evolution of the impact-produced
cloud, the location of the impact could be at 0.2 or 0.32 au (see
Section 4.4). To the zeroth order, we can analytically estimate the
collisional timescale of vapor condensates in such an
environment. When we use equation (13) from \citet{wyatt07}, the
collisional timescale is 0.8--4 yr for 1 cm condensates located at
$r\sim$0.2--0.32 au assuming the condensates are distributed in an
annulus with a width of 1\% of the peak location ($\Delta r/r \sim
 0.01$) and a total mass of $\sim$1$\times10^{-5}M_{\oplus}$. Because the
warm {\it Spitzer} observations only trace small grains, we do not
have direct constraints on the initial mass and volume of the
impact-produced fragments. Therefore, the analytical calculation only
provides some degree of sanity check.

We can also roughly estimate the minimum mass required for the flux
decay due to collisional cascades in such an impact-produced
cloud. When we assume that the dust we detect is at 750 K, the flux difference
(from the highest to the background level is $\sim$2 mJy at 4.5
$\mu$m) suggests a change of 6.4$\times10^{23}$ cm$^2$ in total cross
section for the P1121 system (at 459 pc). If the decrease of
dust cross section is due to collisional cascades from dust grains of
0.5 $\mu$m to 1 mm (grains larger than this size are not probed by 4.5
$\mu$m flux) in a typical power-law size distribution (an index of
--3.5), the total dust mass responsible for this total cross-section
loss is 8$\times10^{-5}M_{Moon}$ ($\sim$10$^{-6}M_{\oplus}$),
equivalent to two $\sim$100 km-size bodies.  We emphasize that these
values should be treated as lower limits because we can
only observe a certain fraction of the dust cross section because of
the optical thickness effects. No additional flux
increase is observed in P1121 (a buildup phase due to the accompanied
boulder population produced in the same impact), therefore the impact product is
most likely dominated by vapor condensates (i.e., a hypervelocity impact). 
In summary, the year-long
flux decay behavior in the P1121 system is consistent with the
aftermath of a hypervelocity impact that produced fragments mostly in
the vapor form.

\begin{figure*} 
  \figurenum{13}
  \label{fig_longterm_flux_evo_id8} 
  \epsscale{1.15} 
  \plotone{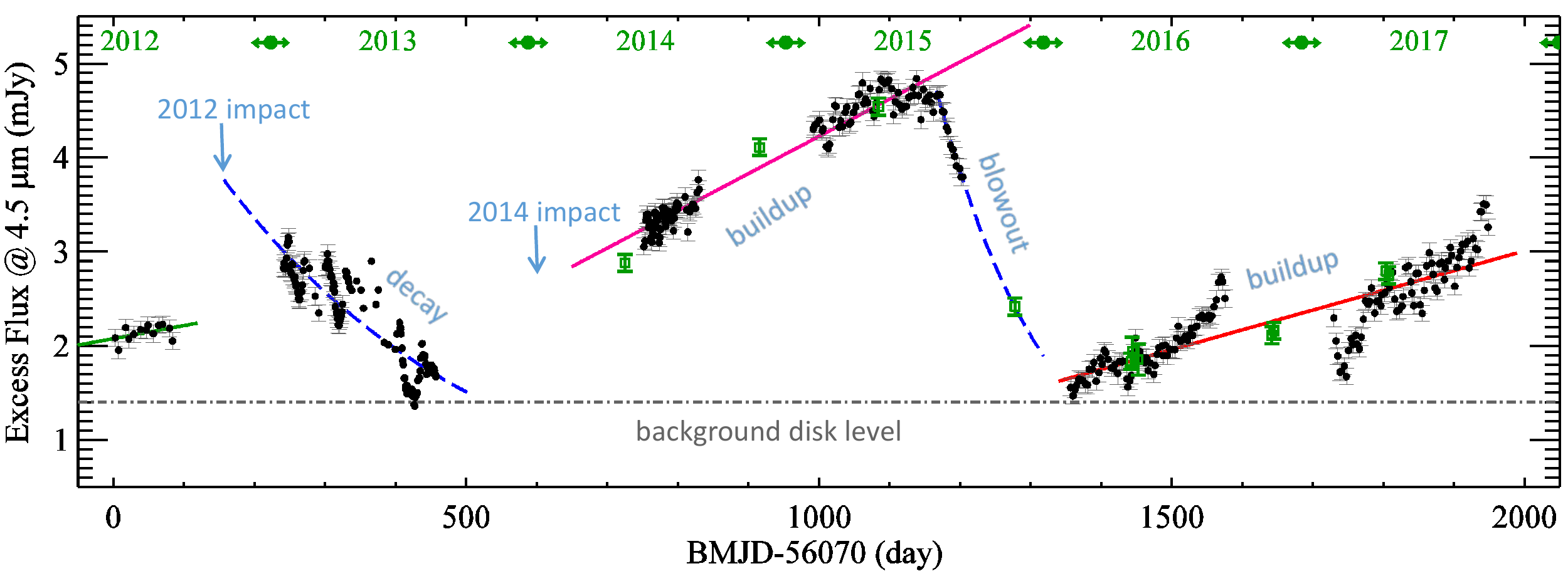}
  \caption{The 4.5 $\mu$m disk flux (dots from {\it Spitzer} and
squares from {\it WISE}) for the ID8 system in 2012-2017 (the year boundaries are
showed at the top of the plot). Various lines show the long-term
trends of the disk variability; the blue dashed curves
represent the exponential fits ($t_0\sim$376 days for the 2013
decay and $t_0\sim$166 days for the 2015 decay), and the linear lines
represent the long-term flux trends. The horizontal dot-dashed line
represents the the maximum level of the background disk emission that
can be identified with the current data. }
\end{figure*}

\subsubsection{Two Large Impacts in the ID8 system}
\label{finalwords_id8}

Given the complex behavior of the disk light curve in the ID8 system, it is possible
that not all variable phenomena are impact related. There will be a wide variety
of scenarios that can explain some part of the variability in the system, which is 
discussed in Section 6.1. Nonetheless, here we aim to explain all of the
variable behaviors using an impact scenario. 

First, we can estimate the typical collision timescale ($t_c$) of 
equal-sized (radius of $s$) planetesimals that are distributed in an annulus
with a distance $r$ from a star using the mean free path estimate. 
Assuming the annulus has a width and a scale height of 10\% of the 
location (i.e., $\delta r = \Delta r/r = 0.1$ and $h = H / r = 0.1$), 
\begin{equation}
\begin{split} t_c & = \frac{1}{\sqrt{2} n \sigma v_c}, \\ & = 12.7 \ {\rm yr} \
\Big[\frac{\rho}{3\ \rm g/cm^{3}} \Big]\Big[\frac{s}{\rm km}\Big]\Big[\frac{r}{\rm 0.2\
au}\Big]^3\Big[\frac{M_{\oplus}}{M_d}\Big] \\ & \ \ \ \ \ \ \ \ \ \ \ \ \ \ \Big[\frac{5\
{\rm km/s}} {v_c}\Big] \Big[\frac{\delta r}{0.1}\Big]\Big[\frac{h}{0.1}\Big],
\end{split}
\end{equation} where $M_d$ is the total mass of the planetesimals,
$\rho$ is the density, and $v_c$= 5 km\ s$^{-1}$ is roughly 10\% of
the Keplerian velocity at 0.2 au. The average collision
timescale\footnote{Equation (13) from \citet{wyatt07}, which has a
slightly different assumption for colliding bodies, gives a timescale
of 7 yr, which is very similar to our value. } of kilometer-size bodies is on
the order of 10 yr, i.e., having two or three large impacts within 10 yr
is possible. However, it is evident that the scale of impacts
inferred from Section \ref{codeMresults} involves bodies with sizes of $\sim$100 km, 
which result in a longer collision timescale for a total mass
of 1 $M_{\oplus}$ in planetesimals (a value expected from the solar
system scale). The {\it Kepler} multiplanet systems tell us that the
typical mass budget for exoplanetary systems is much higher, i.e., the
minimum mass of an extrasolar nebula is 10--100 times higher than the
minimum mass of a solar nebula with a spread of two orders of magnitude
\citep{chiang13,raymond14b}. A high starting mass, a condition that
is likely for these extreme debris disks, further shortens the
timescale. This simple estimate of a collision timescale stands.

There are two clear long-term (more than a few months) flux decays
seen in the data during the past five years: (1) a flux decay since
2013 with a timescale of $\sim$one year and (2) a fast flux drop near
the end of 2015 (Figure \ref{fig_longterm_flux_evo_id8}). Two clear
long-term flux increase trends are also seen: (1) from early 2014 to
late 2015, and (2) from early 2016 to late 2017. Although the infrared
output from the ID8 system varies significantly over the past five
years, it appears that a background level of $\sim$1.4 mJy at 4.5
$\mu$m can be identified (the horizontal dot-dashed line in Figure
\ref{fig_longterm_flux_evo_id8}). Given that we do not have continuous
data coverage (due to visibility gaps), this estimated background
level should be treated as the maximum allowable background level. The
baseline emission likely suggests the presence of kilometer-sized
planetesimals that generate a steady background level of dust. Knowing
this baseline level will help to determine the scale and frequency of
impacts that produce dust above this baseline.

Similar to the P1121 system, the year-long flux decay in 2013 for the
ID8 system is consistent with the rapid collisional evolution from a
swarm of vapor condensates. Using analytical formulae, \citet{meng14}
estimate that the year-long flux decline in the 2013 ID8 disk light
curve is consistent with the collisional cascade timescale of grains
that are no more than a few 100 $\mu$m to millimeter in size. The estimated sizes
in the vapor condensates should be considered lower limits because the
effectiveness of radiation pressure removal must be lower for an
optically thick cloud, which is a required condition to observe
modulations in the light curve.  Similarly, we can also roughly
estimate the minimum mass required for the flux decay observed in
2013. We inferred a total flux of 3.8 mJy at 4.5 $\mu$m on d.d.\
157 (the possible impact date) using the derived exponential
function. Given the estimated background level (1.4 mJy at 4.5
$\mu$m), a flux decrease of 2.4 mJy, corresponding to a total cross
section of 4.8$\times10^{23}$ cm$^2$, is derived (assuming 750 K dust
for a system at 360 pc). Converting the surface area into dust mass, a
lower limit in mass of 5.8$\times10^{-5}$ M$_{Moon}$ is derived,
equivalent to two bodies of equal size, $\sim$100 km-size bodies. In summary,
the short-term modulation due to the optical thickness of the
impact-produced cloud (Section \ref{modulations} and
\ref{modulations_p1121}) on top of the flux decay is consistent with a
violent impact occurring in late 2012 and producing a large amount of
millimeter fragments condensed from vapor, which explains both the
short-term (weekly) and long-term (yearly) trends for up to 2014.

The argument that the ID8 light-curve behavior since 2013 requires an
additional impact comes from the fact that a different period of
short-term modulation was seen in the early 2014 light curves. 
This persisted to early 2015 (Section \ref{modulations}). Under this
assumption, the slow rise in the long-term trend in 2014 and 2015
suggests that the collisional evolution is dominated by the boulder
population with very little vapor. Indeed, when we assume that the 2013 flux
decay did reach the background level near d.d.\ 530 (see Figure
\ref{fig_longterm_flux_evo_id8}), the trend of flux increase in the
beginning of 2014 shows some degree of curvature, similar to the
expected collisional evolution of the boulder population. Furthermore,
a lack of micron-sized fragments in this boulder population is also
required to explain the slow rise. Due to the degeneracy between the
minimum size of the fragments and initial collisional volume, we
cannot determine the exact lower limit in the initial size
distribution of the fragments. Nevertheless, the nominal assumption
that the collisional fragments should extend to the radiation blowout
size (i.e., submicron around solar-like stars) does not apply
here. Under this condition (the fragments produced in the 2014 impact
come from the boulder population), the initial buildup phase takes a
few years to reach quasi-static equilibrium (i.e., maximum total cross
section), and is expected to decay at a very slow rate with a
timescale driven by the collisional timescale of the largest
fragments, ranging from $\sim$10$^2$ to 10$^4$ yr for sizes of 10 m--100 km. 
Therefore, the fast drop seen at the end of 2015 cannot be
explained by the nominal collisional evolution of the boulder
population. We note that the general flux trend (over $\sim$100 days)
is relatively flat prior to the 2015 fast drop, implying that the
collision evolution within the cloud might reach some kind of
equilibrium and trigger the fast-drop event.

We consider several possible mechanisms to explain this rapid flux
drop: \\
\noindent\underline{\it Vapor Condensates}. It might be argued that the rapid
flux drop observed near the end of 2015 could also come from the rapid
evolution of new vapor condensates. To see whether a similar cause
could explain the fast flux drop in 2015, we also fit an exponential
decay curve to the last 40 days of the {\it Spitzer } 2015 data with
the additional {\it WISE} point on d.d.\ 1278, and found that the
decay-time constant is 166 days (see Figure
\ref{fig_longterm_flux_evo_id8}). This timescale is about four times
longer than the orbital period at 0.24 au. If the flux drops in 2013
and 2015 have the same origin, i.e., if they are due to the collisional
destruction of freshly condensed sand-sized grains, the size of the
condensates produced in the flux drop in 2015 must be smaller than those that were 
produced in the 2013 event given the shorter decay timescale\footnote{The
analytical collision lifetime (Equation (13) from \citet{wyatt07})
is on the order of $\sim$200 days for a swarm of 100 $\mu$m particles
with a total mass of 1$\times10^{-4}M_{Moon}$ located at 0.24 au with
an annulus of 0.012 au.}. Smaller droplet sizes generally imply a
smaller size of the impactors (see Figure 13 from
\citealt{johnson12}). Some amount of vapor condensates produced by a
hypervelocity impact between two bodies with sizes of $\sim$10 km (part of the
boulder population produced in the 2014 impact) could quickly destroy
the existing boulder population and produce a fast drop. However, this
should always start with a sharp rise in the very beginning when the
new vapor condensates were added to the existing population, which is
not seen prior to the fast drop in 2015. It might be argued that the optical
thickness of the new vapor population might reduce the amount of flux
in the fast-buildup phase. However, such an optically thick cloud of
new condensates should also have produced short-term modulation,
similar to the 2013 light curve, which was not observed in the fast
2015 flux drop. Therefore, it is difficult to explain the fast 
flux drop in 2015 using the rapid collisional evolution of vapor condensates.

\noindent\underline{\it Coronal Mass Ejection (CME) Event}. 
Alternatively, an energetic CME event, which is expected to be frequent
for a young solar-type star, could also destroy small grains about 
micron size and produce a sudden infrared flux drop, as proposed by
\citet{osten13} to explain the disappearance of dust in TYC8241 26652
1 \citep{melis12}. However, the flux in this case is expected to drop more 
rapidly than what we observed.


\begin{figure*} 
  \figurenum{14}
  \label{movingclump} 
 \epsscale{1.15} 
  \plottwo{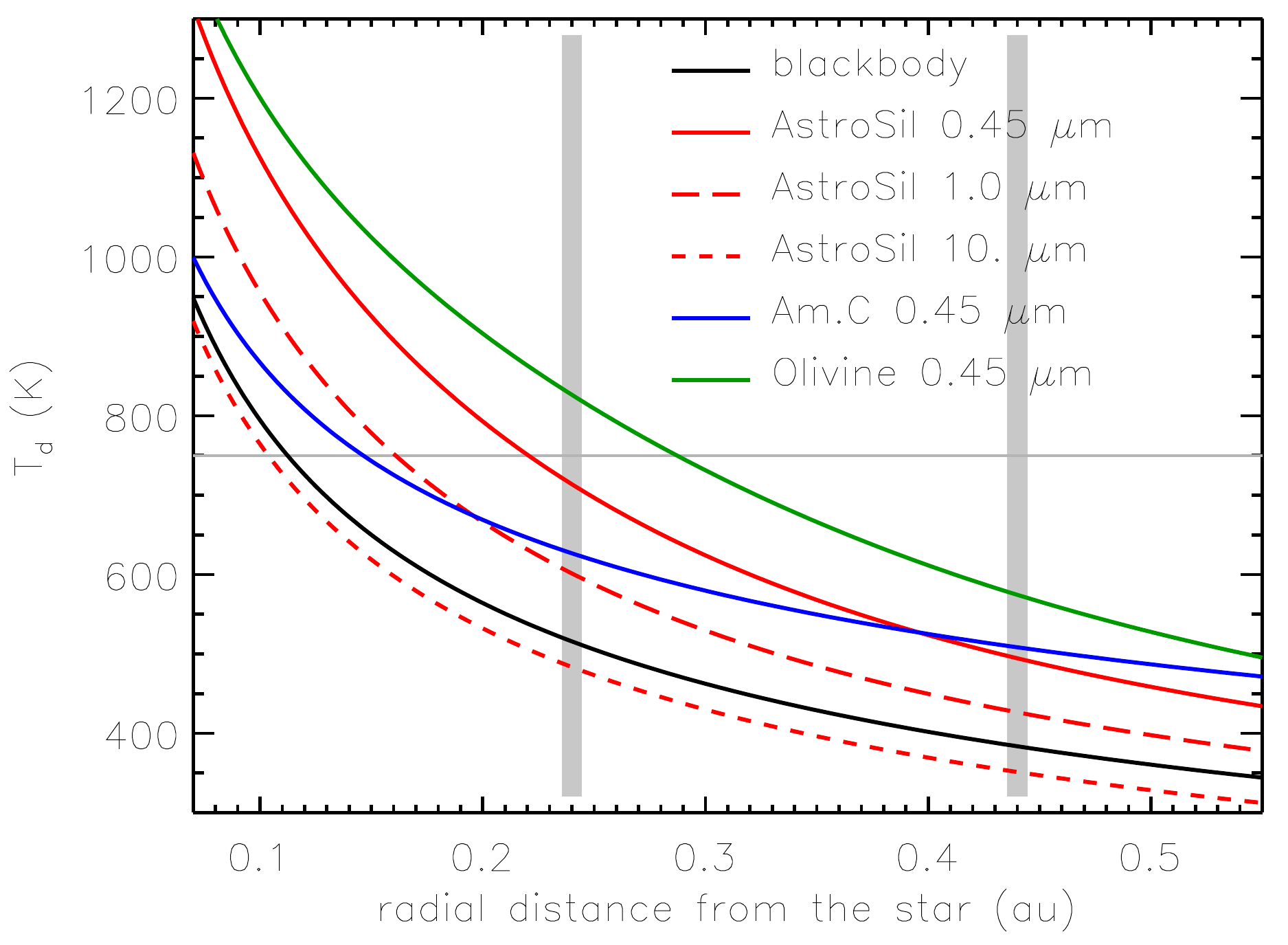}{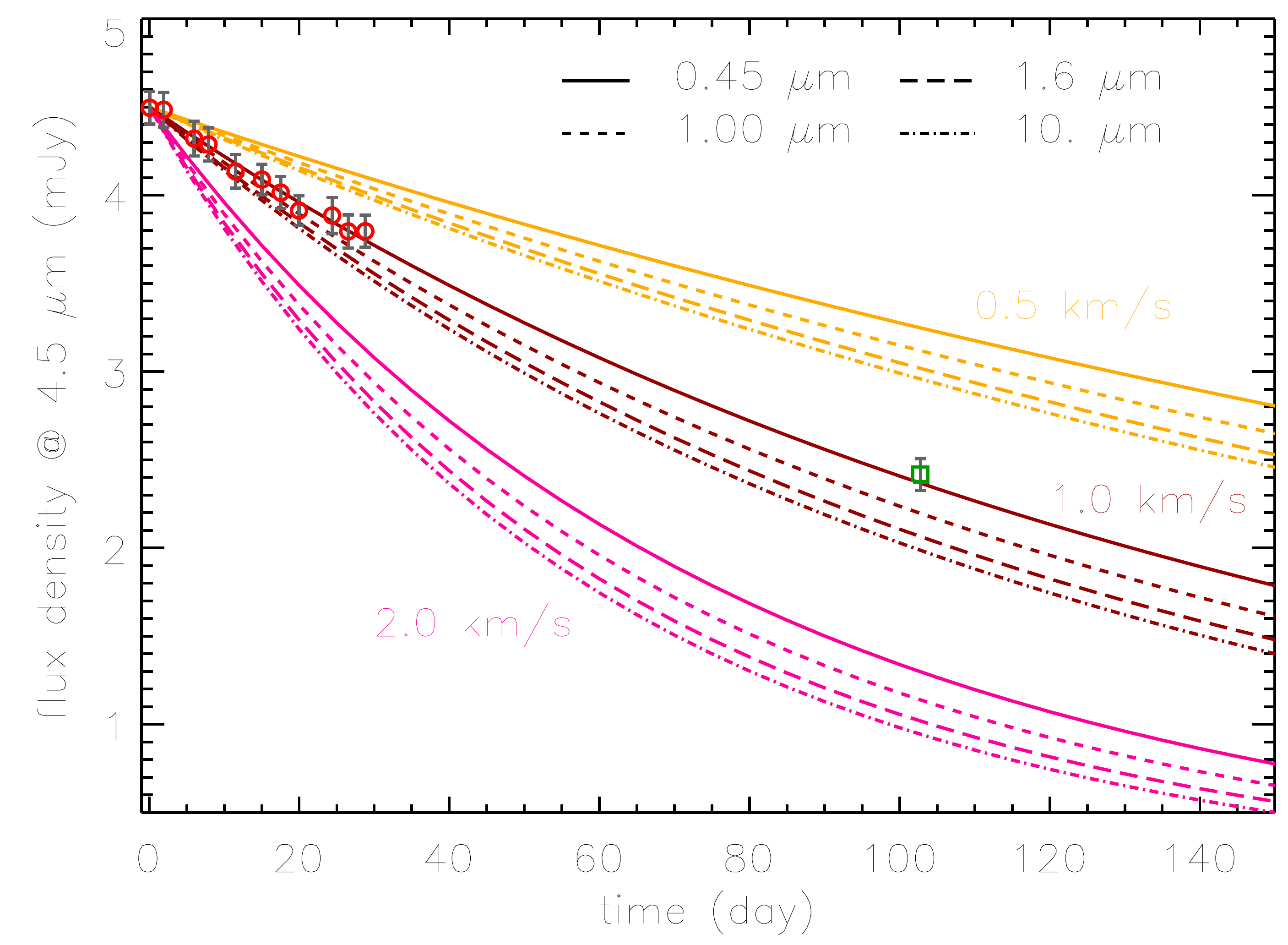}
  \caption{The left panel shows the thermal-equilibrium temperatures in
the ID8 system for a selection of sizes and compositions under the
optically thin assumption. The horizontal line marks the observed
color temperature, 750 K. The two vertical gray lines mark the radial
location of the impacts inferred from the observed short-term
modulations. The right panel shows the expected 4.5 $\mu$m flux evolution
for an optically thin cloud of dust moving outward at three different,
constant velocities (orange for 0.5 km~s$^{-1}$, brown for 1.0
km~s$^{-1}$, and pink for 2.0 km~s$^{-1}$. The cloud is assumed to be
composed of single-size astronomical silicate grains as shown by different styles of
lines. }
\end{figure*}

\noindent\underline{\it Radiation Pressure and Stellar Wind}. Another
fast grain-loss mechanism in a debris system is blowout by radiation
pressure and enhanced stellar wind (proton pressure), with a loss
timescale within a few orbital periods for grains smaller than the
blowout size. To test whether radiation pressure could be related to
the fast flux drop, we compute an expected flux decline due to the
outward motion of a dust cloud. Because the star is the only heating
source in a debris system, the dust temperature, hence the resultant
emission, from a clump of dust, decreases as the clump moves farther
away from the star. As a result, the observed flux decrease is a
delicate balance between the radial velocity of the cloud and where on
the radial temperature profile the grains dominating the emission
lie. Figure \ref{movingclump}a shows the radial dust temperature
profiles in the ID8 system for various grain sizes and
compositions\footnote{We used the optical constants from
\citet{laor93} for astronomical silicates (AstroSil), the constants from
\citet{zubko96} for amorphous carbon (Am.C.), and those from
\citet{dorschner95} for olivine.} under the optically thin
condition. At the radial distance of interest, the dust temperature
profiles are similar for the same composition, with shifts toward lower
temperatures as the grains become larger. Submicron silicate-like
(astronomical silicate or olivine) grains reach a temperature of
$\sim$750 K at 0.24 au. We can then compute the expected flux at 4.5
$\mu$m for a cloud of dust moving radially with a constant
velocity. Although we refer to the dust as a ``cloud'', it is
elongated along its orbit due to Keplerian shear and is not a compact dust
clump. For simplicity, we assume that the cloud is made up of single-size
dust with an optically thin thermal-equilibrium temperature profile,
and has a total mass that yields a total 4.5 $\mu$m flux of 4.5 mJy at
0.24 au initially (time zero). Figure \ref{movingclump}b shows the
flux evolution for different combinations of grain sizes and
velocities. We found that a cloud of 0.45 $\mu$m silicate grains with
a total mass of 1.1$\times10^{21}$ g (1.5$\times10^{-5}$M$_{Moon}$) at
a velocity of 1.0 km~s$^{-1}$ can reproduce the observed flux decay
observed by {\it Spitzer} and {\it WISE} (Figure
\ref{movingclump}b). Although the flux evolution from such a cloud can
fit the 4.5 $\mu$m flux relatively well, the same cloud emission at
3.6 $\mu$m is slightly overpredicted compared with the observed
points (by 2$\sigma$) for the first 20 days, suggesting that the
temperature derived under the optically thin assumption is slightly
too hot, i.e., the cloud might be slightly optically thick
initially. Using the derived velocity of 1.0 km~s$^{-1}$ as the
terminal velocity ($v_r$) for dust being ejected by radiation
pressure, the "effective" $\beta$-value ($\beta$, the ratio between
the radiation pressure and gravitational forces) for the cloud is
0.5002, assuming $v_r \sim v_K \sqrt{2(\beta -0.5)}$ \citep{su05},
where $v_K$ is the Keplerian velocity (58.7 km~s$^{-1}$ at 0.24 au),
consistent with being on an unbound orbit.

Note that we assume a grain density of 2.2 g~cm$^{-3}$ for the above
mass calculation, so the $\beta$ value is 0.5 for the 0.45 $\mu$m
grains (assuming 0.92 M$_{\odot}$ and 0.8 L$_{\odot}$ for the star).
The nominal $\beta$ values for 1 and 10 $\mu$m grains are 0.225 and
0.023, respectively; i.e., they are gravitationally bound to the
system under the minimal condition. For a young solar-type star, the
stellar wind pressure can result in a higher effective $\beta$ value,
$\beta'= (1+ \frac{F_{sw}}{F_{rp}})\beta$ where $F_{sw}$ and $F_{rp}$
are the stellar wind pressure and radiation pressure forces
\citep{burns79}. Large 10 $\mu$m particles can be ejected from the
system when the stellar wind pressure is more than 20 times the
stellar radiation pressure, an enhanced stellar wind phase\footnote{
\citealt{reidemeister11} gives $\frac{F_{sw}}{F_{rp}} =
\frac{\dot{M_{\star}} v_{sw} c}{L_{\star}} \frac{Q_{sw}}{Q_{rp}}$,
where $\dot{M_{\star}}$ is the mass-loss rate, $v_{sw}$ is the wind
velocity, $c$ is the speed of light, $L_{\star}$ is the stellar
luminosity, and $Q_{sw}$ and $Q_{rp}$ are the efficiency factors for
stellar wind and radiation pressure (assumed to be 1). Adopting a
mass-loss rate of 2$\times10^{-10} M_{\sun} yr^{-1}$ for a young
solar-type star during a CME event \citep{cranmer17} and a wind
velocity of 2000 km~s$^{-1}$, this ratio is $\sim$20 during the
enhanced stellar wind phase.}. Even if the large grains could be
ejected during such a phase, the temperatures of such grains are too
low to produce the observed color temperature of $\sim$750 K at the
distance of interest. If the observed flux drop were due to the outward
motion of a dust cloud under the influence of stellar radiation and
wind, the dominant grains in the cloud would have to be about micron to 
submicron size. In summary, the blowout by stellar radiation pressure and
stellar wind is a likely mechanism to explain the 2015 rapid flux
drop.

The remaining question would be how such a large amount
($\sim$$10^{21}$ g) of small dust might suddenly be created among the colliding bodies.
As discussed earlier, the cloud of fragments created by the 2014
impact event is optically thick, and the stellar photons/protons
cannot easily penetrate the innermost part of the cloud, likely
resulting in an overproduction of small grains that should have been
removed. As the cloud becomes sheared by its Keplerian motion and
spread to cover a larger portion of its orbit, the cloud could become
less optically thick and eventually optically thin, a condition at
which the stellar radiation pressure can effectively remove the small
grains. Under this scenario, the 2015 fast flux drop marked the
special time when the impact-produced cloud became somewhat optically
thin (as it has been Keplerian sheared over 2 yr, i.e., more than
15 orbital periods). Being suddenly exposed to the stellar radiation
pressure (and stellar wind) for the first time, the overdense small
dust grains (called seeds by \citet{chiang17}) would
collide with other particles within the cloud to generate grains of
similar sizes ($\beta$-meteoroids) in an exponentially amplifying
avalanche (\citealt{chiang17}; see further discussion in Section
\ref{dustavalanche}). This would change the equilibrium shape of the
small end of the size distribution \citep{wyatt11}, potentially
resulting in a deficit of millimeter-sized grains in the remaining
boulder population. We call this the "reset" boulder population due to
the lack of millimeter sized grains, similar to the condition when these
boulders were first created in 2014.

This might explain the second slow buildup seen in the 2016/2017 light
curve. To illustrate the main difference in the buildup phases
between 2014/2015 and 2016/2017, we show two solid lines in Figure
\ref{fig_longterm_flux_evo_id8} to represent the slope of flux
increase with the same duration of 650 days. Two characteristics are
revealed. First, the buildup phase is not periodic (i.e., the buildup
phase in 2014/2015 is shorter than that of 2016/2017). Second, the
flux increase rate (the slopes in Table 3) in 2016/2017 is shallower
than in 2014/2015. This is consistent with the scenario that we are
seeing the collisional evolution from the same boulder population
created by the 2014 impact, but in a less dense environment (due to
further Keplerian shear). Furthermore, the peak flux in the buildup of
2016/2017 should be lower than the flux in 2015, which is the case
based on the available data. In summary, all the variable behaviors
since the end of 2013 can be explained by one single violent impact
occurring in early 2014.

Admittedly, there is no strong evidence to tie the variable behavior
in the 2014/2015 to the behavior in 2016/2017 because the short-term
modulation seen in the beginning of 2014 disappeared in mid-2015. As
mentioned in Section \ref{modulations}, the only possible dip in 2016
that could be associated with the short-term modulation is on 
d.d.\ 1438, possibly due to one of the disk ansae (orbital phase of
16.75).  Therefore, the slow rise in 2016/2017 could come from the
boulder population created by the 2012 impact. Analogous to the
vapor+boulder A model in Section \ref{codeMresults}, the 2013 decay is
related to the vapor population and the slow rise in 2016/2017 is
related to the boulder population, with both populations generated by
the hypervelocity impact in late 2012.

In conclusion, we need at least two violent impacts to explain all the
variable behaviors: one in late 2012 (at 0.43 au), and the other one in
early 2014 (at 0.24 au), and the slow rise in 2016/2017 could be
associated with either impact. We note that these two impacts might be
related if the second impact was induced by the first,
e.g., if the first impact is some kind of grazing or hit-and-run
collision. It has been shown that the fragments created by a
hit-and-run type of giant impact are likely to return at a later
time and create a secondary impact at a similar location, a possible
scenario that has been proposed to form Mercury
\citep{asphaug14,chau18}. The angle between the collision point and
disk ansa is very similar between the two impact events
($\sim$70\arcdeg\ for the 2012 impact and 90\arcdeg\ for the 2014 one,
derived in Section \ref{modulations}), further supporting this
possibility.

When we relax the assumption that the short-term modulation (10.4 days)
seen in the 2014 and early 2015 light curves is related to the orbital
motion of an impact-produced cloud (i.e., the cloud of fragments did
not need to be at 0.24 au), the slow rise seen in 2014/2015 could be
related to the 2013 violent event that produced both vapor and boulder
populations at 0.43 au, and the 2014/2015 rise could come from the buildup
of the boulder population. By early 2014, this boulder population
would have experienced more than five orbits of Keplerian shearing, and
although still optically thick mostly, produced a minimal short-term
modulation, explaining why we did not see the signature of intermixed
short-term periodicity as observed in the vapor population. Similar to
the scenario described above, the rest of the variable behaviors could
be explained by the same evolution: first, a buildup in the boulder
population due to lack of micron- to millimeter-sized grains initially,
then a rapid 2015 flux drop due to the change from optically thick to
mostly optically thin (a sudden exposure of overdense small grains),
and a second buildup from the "reset" boulder
population. Qualitatively, one single violent impact occurring at the
end of 2012 might explain all the variable behavior except for the
short-term modulation in 2014 and early 2015. Although there are
other mechanisms that might produce the short-term flux modulation,
the modulation amplitude and its disappearing nature are difficult to
explain with non-impact scenarios (see discussion in Section
\ref{otherscenarios}). Therefore, we favor two violent impacts
involving large asteroid-sized bodies to explain the ID8 system, and
suggest that the two impacts might be related as a result of a
grazing or hit-and-run type of impacts.

\section{Discussion }      
\label{discussion}

\subsection{Other Non-impact Scenarios}
\label{otherscenarios}

We have presented a coherent scenario to explain the short- and
long-term (Sections \ref{interpretation_short-term} and
\ref{interpretation_long-term}) disk variability seen in ID8 and
P1121 using the impacts among large asteroid-size bodies. Although we
favor that all the disk variability seen in both systems is impact
related, it is possible that non-impact related scenarios can explain
part of the disk variability given the complex behavior observed by
{\it Spitzer}. One of the puzzling behaviors is the longevity (over 5
yr) of the short-term modulation (a period of 16.7 days and an
amplitude of $\pm$0.08 mJy at 4.5$\mu$m, Figures 5) in the P1121
system.  Although the long-term flux decay (Figure 4) could come from
the collisional evolution of vapor condensates produced by a
hypervelocity impact that occurred prior to 2012, the short-term
modulation due to the optically thick cloud should have been less
notable and disappeared after $\gtrsim$10 orbital periods due to
Keplerian shearing.

We use the short-term modulation observed in the P1121 system as an
example to discuss other non-impact scenarios. Although noisy, the
``flattened'' (after subtracting off the decay trend) excess fluxes
between 3.6 and 4.5 $\mu$m are roughly equal, suggesting a flux ratio
of $\sim$1, i.e., $\sim$1200 K color temperature. Using this as the
dust temperature, the modulation amplitude of $\pm$0.08 mJy at 4.5
$\mu$m for a system at 459 pc suggests an emitting area of
9.84$\times10^{21}$ cm$^2$ (the surface area of an 8 $R_J$ object)
assuming blackbody emission. The required emitting area would be
larger if the dust temperature were lower. Producing the observed
short-term modulation requires a change of such area every 16.7
days. This rules out the possibility of a magma ocean world or a
``Synestia''-like object, a new type of object (with a radius
$\lesssim$$R_J$) formed from a high-energy and angular momentum giant
impact as proposed by \citet{lock17}. The required emitting area
strongly suggests for a form of circumplanetary disk. In fact, an
axisymmetric opaque structure would have a constant projected area as
it rotates and/or orbits around the star with a fixed rotation angle
with respect to the line of sight. To produce some kind of periodic
change in the projected area, the extended disk needs to precess at a
very high rate (like a gyroscope) to produce the modulation of 16.7
days, which is hard to conceive. Similar requirements for the
short-term modulations seen in the ID8 system also apply. Furthermore,
the two different kinds of short-term modulations and their
disappearing nature (fade over a certain period) in ID8 make this
circumplanetary disk scenario unlikely.

There are many ways to deposit new dust into an inner planetary system,
such as collisions in a dense, active asteroid belt \citep{dewit13},
breakups of "super" comets \citep{beichman05}, and volcanic activities
from a newly formed molten planet. Our proposed impact scenario for
both ID8 and P1121 is related to dense, inner planetesimal belts,
the first hypothesis. The other two hypotheses: an Io-like planet or a
comet-like body could undergo repeated bursts of activity,
resulting in a rise in the infrared flux while the eruption/activity
is ongoing, and a decay when the eruption shuts off with a decay on
the lost timescale of the largest produced dust grains. The dust grains
produced in such events would most likely be small, producing the
rapid drop in infrared flux (similar to the collision evolution of
vapor condensates). The rate of dust production from each eruption
would probably vary slightly, as would the size of the largest produced dust
grains, so having slightly different slopes each time would
not be surprising. Furthermore, the repeated nature can be applied
to both the short- and long-term variability.

Another variation of the eruption/activity hypothesis is that the 
dust-producing rate is more or less constant over a multiyear period, but
the loss mechanism is enhanced by the variable nature of the stellar
wind, similar to what has been proposed to explain the fast-moving,
ripple-like features in AU Mic \citep{chiang17}. Because young ($<$150
Myr) solar-like stars do rotate fast with periods ranging from $\sim$1
to a few 10 days (\citealt{gallet13}; a rotation period of $\sim$5 days is
confirmed for ID8, Section \ref{optical}), and their magnetic cycles
(i.e., driven wind) are strongly coupled with stellar rotation
\citep{zanni13}, it is possible to have semi-periodic enhanced
stellar winds to destroy dust. A combination of the repeated
bursts/activities and the variable magnetically driven stellar winds
might explain the complex variable behaviors observed in the extreme
debris systems.  In summary, all the discussed scenarios might
potentially explain some part or all the disk variability, although
the details need further investigation.

\subsection{Minimum Particle Size in an Impact-produced Cloud}

Observing a slow buildup of the infrared excess immediately 
after a violent impact requires that the initial particle size
distribution in the impact-produced swarm has a minimum size on the
order of millimeters or centimeters, as discussed in Section
\ref{codeMresults}. Here, we discuss
possible explanations for the required deficiency in small grains
resulting from a violent impact.

\subsubsection{Condensation from vapor}

The impact locations inferred from the short-term modulations in both
systems range from $\sim$0.2--0.4 au from the star with corresponding
Keplerian velocities of $\sim$40--60 km\ s$^{-1}$. These conditions 
for the debris-producing impact(s) are reminiscent of
conditions near the orbit of Mercury in the early solar system; i.e.,
they occur at a similar distance from a star of similar mass and
luminosity, and the planetesimal mass involved may be within an order
of magnitude of that in the proto-Mercury. The process forming Mercury
itself has proven challenging to understand, but the effort has
produced models for impacts that are directly applicable in similar
situations \citep{benz07, asphaug14, chau18}. The high orbital
velocities lead to extremely violent collisions, and the high
temperatures of the planetesimals due to their proximity to the star
strongly affect the form of the ejecta and their evolution.  The
simulations agree that a large fraction of the mass in solid bodies
can be converted into vapor and then recondensed. Based on the head-on
collision scenario proposed by \citet{benz07} for the Mercury
formation, the condensate particles have a peak in their size
distribution near 1 cm and a rapid drop in the number of particles
below this peak. 

\citet{johnson12} numerically modeled the formation
of vapor condensates for hypervelocity impacts, and found that the
size distribution of the vapor condensates is strongly peaked around
the average value, which depends on the size of impactor and impact
velocity. Therefore, condensation from vapor should result in an upper
limit on the particle size in an impact-produced swarm if the vapor
population is the dominant outcome. However, to reproduce the flux
from the cascade-produced dust requires a high density of particles,
and hence the evolution of the infrared excess proceeds very fast, as
presented in Section \ref{codeMresults}.  In a high-density
environment, the infrared flux evolution of the condensate swarm is
mostly in the decay phase because the initial buildup phase is very
short, which is somewhat independent of the condensate sizes (i.e.,
the evolution is very similar between condensates of 1--5 mm and
condensates of 1--10 cm). Therefore, the slow rise in the 2014 ID8
light curve is unlikely to arise from a swarm of pure vapor
condensates or a ``mixture'' of vapor and boulders.

\subsubsection{Breakup of consolidated bodies}

The distribution of particle sizes yielded by a catastrophic collision
involving consolidated bodies has been studied in a number of
laboratory experiments
\citep[e.g.,][]{davis90,giacomuzzo07,morris17}. The results suggest
that a rollover toward small particles is common with a mass ratio of
10$^{-3}$ to 10$^{-4}$ relative to the target mass, although there is
a scatter of behavior and also some level of concern that the rollover
is a result of the difficulty in finding all of the smallest
debris. These experiments give modest support to the possibility that
a lower size limit arises naturally from the breakup of consolidated
bodies. Taking a mass ratio of 10$^{-9}$ as an extreme value, the
small pieces from a catastrophic breakup of a 100 km consolidated body
are likely to $\sim$100 m. It seems unlikely that the initial size
distribution of the impact-produced swarm of fragments extends to
micron size.

Based on the hypervelocity impact experiments on different types of
target material, \citet{giacomuzzo07} conclude that the fragmentation
is not only governed by impact energy, but that the physical properties of
the bodies (shape and porosity) also play an important role. The growth of
asteroids larger than $\sim$100 km was thought to result from the
accretion of pebbles in a gas-drag-assisted environment
\citep{johansen15}. In this case, a breakup of a large asteroid is
likely to be deficient in grains smaller than the characteristic
``pebble'' size, again, we do not expect to have fragments extending to
micron size.

\subsubsection{Ejected regolith}

The colliding bodies are likely to have accumulated regoliths of small
particles as a result of reaccretion following the many collisions as
they evolved and grew. Some fraction of these regoliths is likely to
be freed in collisions without experiencing the extreme conditions in
the models discussed above, which start from an assumption of
consolidated bodies. These regoliths will also be deficient in small
particles, in this case, because radiation pressure force on small
particles released in an impact will compete with the gravitational
force that reassembles the collision products, and particles that are
sufficiently far from the center of mass of the fragments may be
deflected into orbits that escape the reassembled body. Thus, at
every stage in the assembly of large bodies, the small regolith
particles will tend to be lost from the composite planetesimal.

\citet{burns79}, hereafter B79, explain that the main effect of
radiation pressure on planet-centric dust orbits is to vary the
eccentricity at fixed semimajor axis, with loss when the eccentricity
reaches unity. They show that a loss of dust is expected when (their
Eq.\ 45 and surrounding discussion)

\begin{equation} \beta \sim \frac{0.5\ \mu {\rm m}}{s\rho_1}
\frac{L_*/M_*}{L_\odot/M_\odot} \geq \beta_c \sim
\frac{1}{3}\frac{v}{v_*}
\label{eqno1}
\end{equation}

\noindent where $\beta$ is the ratio of stellar radiation pressure
force to gravitational force on the grain, $s$ is the grain size in
$\mu$m, $\rho_1$ is the ratio of grain density to 3 $g ~ cm^{-3}$
(i.e, dimensionless), and $v$ and $v_*$ are the orbital speeds of the
dust around the planetesimal and the planetesimal around the star.
When circular orbit speeds are used for both $v$ and $v_*$, the loss criterion
becomes

\begin{equation} \beta \geq \left( \frac{M_p}{M_*}\right)^{1/3}
\sqrt{\frac{r_H}{r}}
\label{eqno2}
\end{equation}

\noindent where $M_p$ and $M_\ast$ are the masses of the planetesimal
and the star, $r$ is the distance of the dust from the planetesimal,
and $r_H = a \left(\frac{M_p}{3 M_*} \right)^{1/3}$ is the Hill
radius, where $a$ is the semimajor axis of the orbit of the
planetesimal around the star. Combining equations (\ref{eqno1}) and
(\ref{eqno2}), the critical condition is

\begin{equation} \frac{0.5 \mu m}{s\rho_1} < 2
\frac{L_\ast/M_\ast}{L_\odot/M_\odot} \sqrt{\frac{M_\ast}{M_p}}
\sqrt{\frac{r}{a}}
\label{eqno3}
\end{equation}

\noindent
For the parameters appropriate for ID8 ($a$ = 0.24 au, $L_\ast/M_\ast
\sim L_\odot/M_\odot$), small planetesimals will lose centimeter-sized
fragments due to this mechanism in the regolith.  For example, an object with 
a diameter of 100 m will lose fragments $<$40 cm in size that are more
than 10 m from the centers of mass of the parent objects. If the
impact fragments are dominated by the regolith from the colliding
bodies (e.g., less energetic, cratering impact events), this
``characteristic size'' serves as the minimum particle size for the
impact-produced cloud.

These expectations are supported by studies of the thermal inertia of
asteroid regoliths. \citet{gundlach13} deduced from the thermal
inertias that asteroids with diameters smaller than $\sim$100 km have
relatively coarse regolith grains with particles typically in the range of 
millimeters to centimeters range.

In summary, all of the processes that might contribute to the initial
particle size of the impact-produced cloud are potentially subject to
a minimum size limit, consistent with the requirements from the
simulations presented in Section \ref{codeMresults}.

\subsection{Unsolved Issues for the ID8 Rapid Flux Drop in 2015 }
 \label{dustavalanche}

Our basic idea in explaining the rapid flux drop observed in the ID8
system near the end of 2015 is similar to but different from the dust
avalanche calculations as described by \citet{grigorieva07} and
\citet{thebault18}, where some amount of unbound grains released by a
breakup of a planetesimal sandblasts an existing outer planetesimal
belt in which a quasi-static state collisional cascade has been
established. In our proposed scenario, the initial ``unbound'' grains
are accumulating in a confined region where stellar photons/protons
cannot easily penetrate due to the thickness of an impact-produced
cloud. In such an environment, collisional cascades would create
grains that are smaller than the nominal blowout size without being
blown out, and eventually small enough that radiation pressure cannot
remove them at all. This might be a mechanism to create abundant,
stable submicron grains in the HD172555 system, as proposed by
\citet{johnson12b}. 


\citet{thebault18} revisited the dust avalanche calculations as
originally proposed by \citet{grigorieva07}, and concluded that (1)
the avalanches are more effective at the close-in location because the
velocity of the unbound grains, an important factor for catastrophic
collisions, is proportional to the Keplerian velocity, and (2) the
duration of the avalanche (in both the increase and decrease flux
phase) is very short, on the order of the fractional orbital
timescale of the planetesimal belt. Both conclusions appear to fit the condition
in the ID8 system. However, the authors further concluded that the luminosity
deficit, as compared to the pre-avalanche level, remains very limited
because the preexisting unbound (native as in the existing
planetesimal belt) grains shield the native small bound grains that
make up most of the target reservoir for fueling the avalanche
propagation. Although the mass of the breakup population (seeds), on
the order of $10^{20-22}$ g in their calculations, is roughly similar
to the mass of the moving clump estimated in Section
\ref{finalwords_id8}, the numbers of the initial unbound grains are
quite different because their seeds have a size distribution extending
to 1 cm when counting the mass, which is roughly two orders of
magnitude lower. As discussed in Section \ref{codeMresults}, the
density of colliding particles is an important factor in determining
the collisional evolution. The structure of the dust avalanche zone
(i.e., the impact-produced cloud in our scenario, or the outer
planetesimal belt in their calculation) and the number of initial
unbound grains are both important factors to determine the efficiency
of collisional avalanches. A future investigation is needed to
further test our proposed scenario.

\section{Conclusions }
\label{conclusion}

We reported warm {\it Spitzer} monitoring data at 3.6 and 4.5 $\mu$m
for ID8 and P1121, taken in the past 5 yr. Time-series observations
with cadences of $\sim$1--3 days were obtained during each object's visibility
windows, supplemented by {\it WISE} data during the visibility
gaps. The extended coverage of the {\it Spitzer} data revealed complex
variable behaviors that can be characterized as short-term (weekly to
month) and long-term (yearly) variability. For the P1121 system, our
new {\it Spitzer} data confirmed the year-long flux decay reported by
\citet{meng15}, and further revealed that the disk flux reached to a
background level since 2015. In addition to the flux decay, Fourier
analysis also revealed a short-term modulation with a period of 16.7$\pm$1.5
days that persisted over the past 5 yr.  For
the ID8 system, our new {\it Spitzer} data revealed dramatically
different behaviors from the behavior reported by \citet{meng14}, where two
intermixed periods (26 and 34 days) were found on top of a general
flux decay using the 2013 data. Instead of flux decrease, the {\it
Spitzer} data in 2014/2015 and 2016/2017 showed long-term flux
increases with a sharp drop near the end of 2015 visibility window. A
maximum allowable background level ($\sim$1.4 mJy at 4.5 $\mu$m) was
also identified. Furthermore, a new flux modulation with a single
period of 10.4$\pm$1.5 days was discovered in addition to the 2014/2015 flux
increase, but became less noticeable in mid-2015.

We obtained ground-based optical photometry for both systems during
the {\it Spitzer} visibility windows whenever possible, for which we used a robotic
telescope in Chile. Combining our results with the data obtained from the KELT and
ASAS-SN networks, we found that both stars are stable in the optical
within a few hundredth magnitude. For ID8, we confirmed a weak (0.01 mag in
$V$ band) modulation with a period of 5 days that is due to spots on the
stellar surface, for which we used more than 5 yr of the optical data. This
suggests that the rotation axis of the star is inclined from the line of
sight. No significant optical periodicity was found for P1121,
suggesting the system might be viewed close to pole-on. We also
inferred the debris location using SED models to fit the mid-infrared
spectrum of P1121, which shows prominent silicate features in the 10
$\mu$m region. Similar to the ID8 system \citep{olofsson12}, the
results favor a close-in location, ranging from 0.2 to 1.6 au, with a
total dust mass of 9.0$\times10^{-6} M_{\oplus}$ (up to 1 mm).

We posit that the complex infrared variability in both systems can
be explained by one single hypothesis -- the aftermaths of violent
impacts. Debris generated by a violent impact forms a thick cloud of
fragments, which is further sheared by its Keplerian motion as it
orbits around the star. Under the dynamical evolution
\citep{jackson14}, the projected area of such a cloud reaches local
minima at the collision and anti-collision points (i.e.,
bi-periodicity), and possibly with two additional minima along the
disk ansae if the system is viewed close to edge-on
\citep{meng14,jackson19}. These two bi-periodicity effects are
independent of each other; the phase difference only depends
on the relative orbital locations between where the impact occurs and
the disk ansae. The impact debris is characterized by a mixture of
vapor condensates (about millimeter to centimeter size) and escaping boulders, with a
relative ratio of the two depending on the impact conditions. In a
hypervelocity impact, the energy involved could totally vaporize the
impactor, i.e., vapor condensates are the dominant impact product. The
impact-generated fragments, once released, would start to collide among
themselves and with any existing debris to generate fine dust that
emits efficiently in the infrared. The combination of the dynamical
and collisional evolution from an impact-produced cloud produces the
complex infrared variability that can be monitored by infrared
observations. Given the large range of particle sizes involved in such
an impact-produced cloud, it is numerically challenging to couple the
dynamical and collisional evolution of the cloud self-consistently. 
We therefore qualitatively modeled the short-term and
long-term variability separately using existing codes to extract
basic parameters about the impacts.

Using 3D radiative transfer calculations, we demonstrated that an
impact-produced clump of optically thick dust, under the influence of
the dynamical and viewing geometric effects \citep{jackson14,jackson19}, can
produce short-term modulation in the disk light curves. Right after an
impact, the lowest fluxes (dips) always occur at the collision and
anti-collision points, and the second lowest fluxes occur at the disk ansae
for an inclined geometry. The times at which the dips occur can be
used to determine the true orbital period of the impact-produced
cloud, and the relative phase between the collision point and the disk
ansa for an inclined geometry. Because the infrared observations are
most sensitive to small micron-sized particles, the long-term
evolution of the infrared flux is governed by how fast the small micron-sized 
dust is being generated in the cloud; i.e., it depends
sensitively on the initial size distribution and on the density of the
impact fragments. Using a 1D collisional cascade code, we
demonstrated that the long-term flux trend for a swarm of millimeter to centimeter
vapor condensates exhibits a quick rise-and-fall behavior -- a sharp
increase (i.e., buildup) in infrared flux followed by a flux decay
once the swarm of particles reaches the quasi-static state collisional
cascades and starts to deplete the mass of the largest fragments. The
buildup phase for the boulder population could be long, depending
sensitively on the minimum size of the fragments, i.e., the larger the
minimum size, the longer the buildup phase. The rate of the flux decay
depends on the collisional timescale of the largest fragments in the
swarm. Therefore, it is more likely to observe a flux decay from a
swarm of vapor condensates than from a swarm of large boulders.
Finally, a combination of the two different flux trends (rise-and-fall
from vapor condensate plus a secondary rise from boulders) is expected
if the violent impact produced a ``mixture'' of boulders and vapor
condensates (a typical power-law size distribution of fragments plus
some extra amount of fragments at the smaller size end as vapor
condensates).

Based on our qualitative modeling results, we concluded that the
infrared variability observed in the P1121 system is most likely
resulting from a hypervelocity impact that occurred prior to 2012. The
infrared flux decay with a characteristic timescale of one year suggests
that the impact fragments were dominated by the vapor condensates of
millimeter to centimeter sizes resulting from a very violent (i.e., hypervelocity)
impact. The short-term modulation with a single period of 16.7 days
suggests a true orbital period of 33.4 days (for a face-on geometry
with a semimajor axis of 0.2 au) or 66.8 days (for an edge-on geometry
with a semimajor axis of 0.42 au plus that the collision point was 
halfway between the disk ansae). Because we did not detect any
significant optical modulation due to stellar spots for this 80 Myr
old system, the star is likely to be viewed close to
pole-on. Therefore, a hypervelocity impact occurring at 0.2 au is more
likely if the short-term modulation were due to the orbital motion of
the impact-produced cloud. The argument against such an assumption is
due to the fact the observed short-term modulation lasted for more
than $\sim$30--50 orbits even when the cloud emission had reached the
background level since 2015. We discussed several non-impact
scenarios that might explain the observed short-term modulation in
Section \ref{otherscenarios}. All of them (including the impact
scenario) need further investigation.

For the ID8 system, the 2013 data presented by \citet{meng14} are very
similar to the variable behavior in the P1121 system, except that ID8
is likely to be viewed at an inclined angle, so the short-term
modulation should show additional bi-periodicity due to disk ansae. We
reinterpreted the 2013 light curve and identified the true orbital
period of the impact-produced cloud to be 108 days (i.e., at 0.43
au). The angle between the disk ansa and the collision point was
likely to be $\sim$70\arcdeg, which created the intermixed periodicity
reported by \citet{meng14}. The 2013 short-term modulation on top of a
flux decay is consistent with the dynamical and collisional evolution
from an optically thick cloud of millimeter to centimeter vapor condensates,
generated by a hypervelocity impact in late 2012. The fact that we
detected a new, single, short-term modulation on top of a slow flux
increase in 2014 argues that a new impact occurred during the {\it
Spitzer} visibility gap in early 2014. The single periodicity
suggests that the new collision point occurred at 90\arcdeg\ from the
disk ansa with a true orbital period of 41.6 days (i.e., 0.24
au). This new short-term modulation disappeared (or became less
prominent) in mid-2015 as the impact-produced cloud has been sheared
by $\sim$10 orbits. The initial slow buildup phase in 2014/2015
suggests that the fragments were dominated by boulders with very
little of vapor condensates, and had a minimum size of about millimeters
to centimeters. In the beginning of the post-impact collision evolution,
stellar radiation pressure could only effectively remove the small
grains near the surface of the optically thick cloud, leading to an
overproduction of small grains in the center of the cloud. As the
cloud was continuously being sheared and spread out, the overdense
small grains became more transparent to the stellar photons/protons as
the conditions changed to less optically thick, a runaway effect
quickly destroyed the newly generated small grain population,
producing the rapid flux drop seen in the end of 2015, and resetting
the minimum size of the boulder population to millimeter to centimeter sizes
(again). The second slow buildup seen in the 2016/2017 is consistent
with the ``reset'' boulder population ever since. Alternatively, the
slow rise in 2016/2017 could come from the accompanied boulder
population created by the 2012 impact event that created a mixture of
vapor condensates and boulders. In summary, the observed infrared
variability in the ID8 system in the past 5 yr was consistent with
two violent impact events -- one in late 2012, and the other in early
2014. Because the angles between the collision point to the disk ansa
were similar between the two events, we further suggested that these
two impacts might be related as a result of grazing or hit-and-run
type of events.

Limited by the available data of these two systems, we could not
precisely determine the size of bodies involved in these violent
impacts. However, the changes in the dust cross section (i.e., flux)
due to the collisional cascades in the impact-produced clouds suggests
that bodies of $\gtrsim$100 km were involved. A future self-consistent
numerical model that can track the evolution of all sizes of particles
dynamically and collisionally might extract more information about
these violent events.  Finally, the extraordinary photometry
precision, high cadence, and long-baseline observations provided by
{\it Spitzer} enable detailed documenting of the disk variability in
extreme debris disks, and provide unique observational insights into
the processes of terrestrial planet formation.

\acknowledgments

This work is based on observations made with the {\it Spitzer} Space
Telescope, which is operated by the Jet Propulsion Laboratory,
California Institute of Technology. K.Y.L.S. is grateful for funding
from NASA's ADAP program (grant No.\ NNX17AF03G). G.M.K. is
supported by the Royal Society as a Royal Society University Research
Fellow. R.M. acknowledges funding from NASA Exoplanets Research
Program (grant 80NSSC18K0397). J.O. acknowledges financial support
from the ICM (Iniciativa Cient\'ifica Milenio) via the N\'ucleo
Milenio de Formaci\'on Planetaria grant, from the Universidad de
Valpara\'iso, and from Fondecyt (grant 1180395).

This publication makes use of data products from the Near-Earth Object
Wide-field Infrared Survey Explorer ({\it WISE}), which is a project
of the Jet Propulsion Laboratory/California Institute of
Technology. {\it WISE} is funded by the National Aeronautics and Space
Administration. This work has made use of data from the European Space
Agency (ESA) mission {\it Gaia}
(\url{https://www.cosmos.esa.int/gaia}), processed by the {\it Gaia}
Data Processing and Analysis Consortium (DPAC,
\url{https://www.cosmos.esa.int/web/gaia/dpac/consortium}). Funding
for the DPAC has been provided by national institutions, in particular
the institutions participating in the {\it Gaia} Multilateral
Agreement. 

\facility{Spitzer (IRAC, IRS)}

\onecolumngrid

\begin{sidewaysfigure*}
  \figurenum{1}
  \begin{center}
    \centerline{\includegraphics[width=\linewidth]{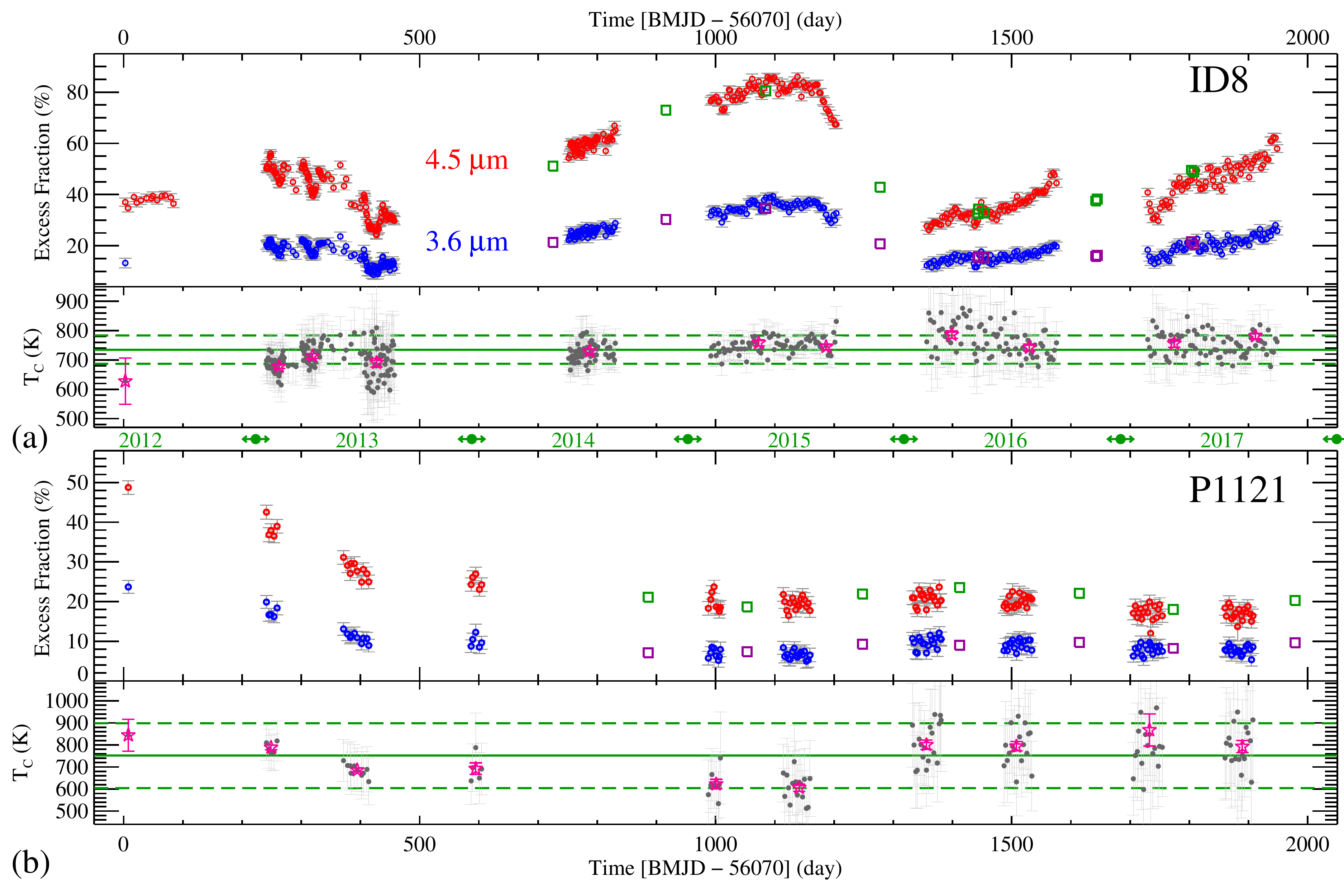}}%
    \label{timeseries_id8_p1121}
    \figcaption{Time-series excesses (upper panel) and corresponding color
temperatures (bottom panel) for the ID8 system in (a) and the P1121
system in (b). For both (a) and (b), the excess fluxes are shown
relative to the stellar photosphere (stable within 1\%) in the upper
panel. The open circles are the {\it Spitzer} measurements, while the
squares are from {\it WISE} (purple for $W1$ and dark green for $W2$ after
photospheric subtraction and small offset adjustments). In the bottom
panel, the star symbols are the time-average (one to a few per
visibility window) color temperatures, and the horizontal solid line
represents the average color temperature over the past 5 yr with
the dashed lines for the $\pm$1 $\sigma$ variation. }
\end{center}
\end{sidewaysfigure*}


\begin{thebibliography}{}

\bibitem[Agnor \& Asphaug(2004)]{agnor04} Agnor, C., \& Asphaug, E.\ 2004, \apjl, 613, L157 

\bibitem[Asphaug \& Reufer (2014)]{asphaug14} Asphaug, E., \& Reufer, A. 2014, Nature Geoscience, 7, 564

\bibitem[Balog et al.(2009)]{balog09} Balog, Z., Kiss, L.~L., Vink{\'o}, J., et al.\ 2009, \apj, 698, 1989 

\bibitem[Beichman et al.(2005)]{beichman05} Beichman, C.~A., Bryden, G., Gautier, T.~N., et al.\ 2005, \apj, 626, 1061 

\bibitem[Benz et al.(1988)]{benz88} Benz, W., Slattery, W.~L., \& Cameron, A.~G.~W.\ 1988, \icarus, 74, 516 

\bibitem[Benz et al.(2007)]{benz07} Benz, W., Anic, A., Horner, J., \& Whitby, J.~A.\ 2007, \ssr, 132, 189 

\bibitem[Booth et al.(2009)]{booth09} Booth, M., Wyatt, M.~C., Morbidelli, A., Moro-Mart{\'{\i}}n, A., \& Levison, H.~F.\ 2009, \mnras, 399, 385 

\bibitem[Bottke \& Norman(2017)]{bottke17} Bottke, W.~F., \& Norman, M.~D.\ 2017, Annual Review of Earth and Planetary Sciences, 45, 619 

\bibitem[Burns et al.(1979)]{burns79} Burns, J.~A., Lamy, P.~L., \& Soter, S.\ 1979, \icarus, 40, 1 

\bibitem[Chambers \& Wetherill(1998)]{chambers98} Chambers, J.~E., \& Wetherill, G.~W.\ 1998, \icarus, 136, 304 

\bibitem[Chambers(2013)]{chambers13} Chambers, J.~E.\ 2013, \icarus, 224, 43 

\bibitem[Chau et al.(2018)]{chau18} Chau, A., Reinhardt, C., Helled, R., \& Stadel, J. 2018, ApJ, 865, 35

\bibitem[Chiang \& Laughlin(2013)]{chiang13} Chiang, E., \& Laughlin, G.\ 2013, \mnras, 431, 3444 

\bibitem[Chiang \& Fung(2017)]{chiang17} Chiang, E., \& Fung, J.\ 2017, \apj, 848, 4 

\bibitem[Cranmer(2017)]{cranmer17} Cranmer, S.~R.\ 2017, \apj, 840, 114 

\bibitem[Davis \& Ryan(1990)]{davis90} Davis, D.~R., \& Ryan, E.~V.\ 1990, \icarus, 83, 156 

\bibitem[de Wit et al.(2013)]{dewit13} de Wit, W.~J., Grinin, V.~P., Potravnov, I.~S., et al.\ 2013, \aap, 553, L1 

\bibitem[Dong et al.(2012)]{dong12} Dong, R., Rafikov, R., Zhu, Z., et al.\ 2012, \apj, 750, 161 

\bibitem[Dong et al.(2015)]{dong15} Dong, R., Zhu, Z., Rafikov, R.~R., \& Stone, J.~M.\ 2015, \apjl, 809, L5 

\bibitem[Dorschner et al.(1995)]{dorschner95} Dorschner, J., Begemann, B., Henning, T., Jaeger, C., \& Mutschke, H.\ 1995, \aap, 300, 503 

\bibitem[Gaia Collaboration et al.(2016)]{gaia16} Gaia Collaboration, Prusti, T., de Bruijne, J.~H.~J., et al.\ 2016, \aap, 595, A1 

\bibitem[Gaia Collaboration et al.(2018)]{gaia18} Gaia Collaboration, Brown, A.~G.~A., Vallenari, A., et al.\ 2018, arXiv:1804.09365 

\bibitem[Gallet \& Bouvier(2013)]{gallet13} Gallet, F., \& Bouvier, J.\ 2013, \aap, 556, A36 

\bibitem[G{\'a}sp{\'a}r et al.(2012a)]{gaspar12} G{\'a}sp{\'a}r, A., Psaltis, D., {\"O}zel, F., Rieke, G.~H., \& Cooney, A.\ 2012, \apj, 749, 14 

\bibitem[Genda et al.(2015)]{genda15} Genda, H., Kobayashi, H., \& Kokubo, E.\ 2015, \apj, 810, 136 

\bibitem[Giacomuzzo et al.(2007)]{giacomuzzo07} Giacomuzzo, C., Ferri, F., Bettella, A., et al.\ 2007, Advances in Space Research, 40, 244

\bibitem[Gorlova et al.(2004)]{gorlova04} Gorlova, N., Padgett, D.~L., Rieke, G.~H., et al.\ 2004, \apjs, 154, 448 

\bibitem[Grigorieva et al.(2007)]{grigorieva07} Grigorieva, A., Artymowicz, P., \& Th{\'e}bault, P.\ 2007, \aap, 461, 537 

\bibitem[Gundlach \& Blum(2013)]{gundlach13} Gundlach, B., \& Blum, J.\ 2013, \icarus, 223, 479 

\bibitem[Jackson \& Wyatt(2012)]{jackson12} Jackson, A.~P., \& Wyatt, M.~C.\ 2012, \mnras, 425, 657 

\bibitem[Jackson et al.(2014)]{jackson14} Jackson, A.~P., Wyatt, M.~C., Bonsor, A., \& Veras, D.\ 2014, \mnras, 440, 3757 

\bibitem[Jackson et al.(2019)]{jackson19} Jackson, A.~P., et al.\ 2019, in prep. 

\bibitem[Jarrett et al.(2011)]{jarrett11} Jarrett, T.~H., Cohen, M., Masci, F., et al.\ 2011, \apj, 735, 112 

\bibitem[Johansen et al.(2015)]{johansen15} Johansen, A., Mac Low, M.-M., Lacerda, P., \& Bizzarro, M.\ 2015, Science Advances, 1, 1500109 

\bibitem[Johns-Krull et al.(2016)]{johns-krull16} Johns-Krull, C.~M., McLane, J.~N., Prato, L., et al.\ 2016, \apj, 826, 206 

\bibitem[Johnson \& Melosh(2012)]{johnson12} Johnson, B.~C., \& Melosh, H.~J.\ 2012, \icarus, 217, 416 

\bibitem[Johnson et al.(2012)]{johnson12b} Johnson, B.~C., Lisse, C.~M., Chen, C.~H., et al.\ 2012, \apj, 761, 45 

\bibitem[Kennedy \& Wyatt(2013)]{kennedy13} Kennedy, G.~M., \& Wyatt, M.~C.\ 2013, \mnras, 433, 2334 

\bibitem[Kennedy et al.(2017)]{kennedy17} Kennedy, G.~M., Kenworthy, M.~A., Pepper, J., et al.\ 2017, Royal Society Open Science, 4, 160652 

\bibitem[Kenyon \& Bromley(2004)]{kenyon04} Kenyon, S.~J., \& Bromley, B.~C.\ 2004, \apjl, 602, L133 

\bibitem[Kenyon \& Bromley(2006)]{kenyon06} Kenyon, S.~J., \& Bromley, B.~C.\ 2006, \aj, 131, 1837 

\bibitem[Kenyon \& Bromley(2016)]{kenyon16} Kenyon, S.~J., \& Bromley, B.~C.\ 2016, \apj, 817, 51 

\bibitem[Kim et al.(1994)]{kim94} Kim, S.-H., Martin, P.~G., \& Hendry, P.~D.\ 1994, \apj, 422, 164 

\bibitem[Kochanek et al.(2017)]{kochanek17} Kochanek, C.~S., Shappee, B.~J., Stanek, K.~Z., et al.\ 2017, \pasp, 129, 104502 

\bibitem[Laor \& Draine(1993)]{laor93} Laor, A., \& Draine, B.~T.\ 1993, \apj, 402, 441 

\bibitem[Lebouteiller et al.(2011)]{lebouteiller11} Lebouteiller, V., Barry, D.~J., Spoon, H.~W.~W., et al.\ 2011, \apjs, 196, 8 

\bibitem[Leinhardt \& Stewart(2012)]{leinhardt12} Leinhardt, Z.~M., \& Stewart, S.~T.\ 2012, \apj, 745, 79 

\bibitem[Lock \& Stewart(2017)]{lock17} Lock, S.~J., \& Stewart, S.~T.\ 2017, Journal of Geophysical Research (Planets), 122, 950 

\bibitem[Lock et al.(2018)]{lock18} Lock, S.~J., Stewart, S.~T., Petaev, M.~I., et al.\ 2018, Journal of Geophysical Research (Planets), 123, 910 

\bibitem[Kral et al.(2015)]{kral15} Kral, Q., Th{\'e}bault, P., Augereau, J.-C., Boccaletti, A., \& Charnoz, S.\ 2015, \aap, 573, A39 

\bibitem[Mainzer et al.(2011)]{mainzer11} Mainzer, A., Bauer, J., Grav, T., et al.\ 2011, \apj, 731, 53 

\bibitem[Mainzer et al.(2014)]{mainzer14} Mainzer, A., Bauer, J., Cutri, R.~M., et al.\ 2014, \apj, 792, 30 

\bibitem[Melis et al.(2010)]{melis10} Melis, C., Zuckerman, B., Rhee, J.~H., \& Song, I.\ 2010, \apjl, 717, L57 

\bibitem[Melis et al.(2012)]{melis12} Melis, C., Zuckerman, B., Rhee, J.~H., et al.\ 2012, \nat, 487, 74 

\bibitem[Meng et al.(2012)]{meng12} Meng, H.~Y.~A., Rieke, G.~H., Su, K.~Y.~L., et al.\ 2012, \apjl, 751, L17 

\bibitem[Meng et al.(2014)]{meng14} Meng, H.~Y.~A., Su, K.~Y.~L., Rieke, G.~H., et al.\ 2014, Science, 345, 1032 

\bibitem[Meng et al.(2015)]{meng15} Meng, H.~Y.~A., Su, K.~Y.~L., Rieke, G.~H., et al.\ 2015, \apj, 805, 77 

\bibitem[Meng et al.(2017)]{meng17} Meng, H.~Y.~A., Rieke, G.~H., Su, K.~Y.~L., \& G{\'a}sp{\'a}r, A.\ 2017, \apj, 836, 34 

\bibitem[Morlok et al.(2014)]{morlok14} Morlok, A., Mason, A.~B., Anand, M., et al.\ 2014, \icarus, 239, 1 

\bibitem[Morris \& Burchell(2017)]{morris17} Morris, A.~J.~W., \& Burchell, M.~J.\ 2017, \icarus, 296, 91 

\bibitem[Mustill \& Wyatt(2009)]{mustill09} Mustill, A.~J., \& Wyatt, M.~C.\ 2009, \mnras, 399, 1403 

\bibitem[Olofsson et al.(2012)]{olofsson12} Olofsson, J., Juh{\'a}sz, A., Henning, T., et al.\ 2012, \aap, 542, A90 

\bibitem[Osten et al.(2013)]{osten13} Osten, R., Livio, M., Lubow, S., et al.\ 2013, \apjl, 765, L44 

\bibitem[Pepper et al.(2007)]{pepper07} Pepper, J., Pogge, R.~W., DePoy, D.~L., et al.\ 2007, \pasp, 119, 923 

\bibitem[Pepper et al.(2012)]{pepper12} Pepper, J., Kuhn, R.~B., Siverd, R., James, D., \& Stassun, K.\ 2012, \pasp, 124, 230 

\bibitem[Raymond et al.(2014)]{raymond14} Raymond, S.~N., Kokubo, E., Morbidelli, A., Morishima, R., \& Walsh, K.~J.\ 2014, Protostars and Planets VI, 595 

\bibitem[Raymond \& Cossou(2014)]{raymond14b} Raymond, S.~N., \& Cossou, C.\ 2014, \mnras, 440, L11 

\bibitem[Reach et al.(2005)]{reach05} Reach, W.~T., Megeath, S.~T., Cohen, M., et al.\ 2005, \pasp, 117, 978 

\bibitem[Rebull et al.(2014)]{rebull14} Rebull, L.~M., Cody, A.~M., Covey, K.~R., et al.\ 2014, \aj, 148, 92 

\bibitem[Reegen(2007)]{reegen07} Reegen, P.\ 2007, \aap, 467, 1353 

\bibitem[Reidemeister et al.(2011)]{reidemeister11} Reidemeister, M., Krivov, A.~V., Stark, C.~C., et al.\ 2011, \aap, 527, A57 

\bibitem[Sallum et al.(2015)]{sallum15} Sallum, S., Follette, K.~B., Eisner, J.~A., et al.\ 2015, \nat, 527, 342 

\bibitem[Shappee et al.(2014)]{shappee14} Shappee, B.~J., Prieto, J.~L., Grupe, D., et al.\ 2014, \apj, 788, 48 

\bibitem[Stellingwerf(1978)]{stellingwerf78} Stellingwerf, R.~F.\ 1978, \apj, 224, 953 

\bibitem[Stewart \& Leinhardt(2012)]{stewart12} Stewart, S.~T., \& Leinhardt, Z.~M.\ 2012, \apj, 751, 32 

\bibitem[Su et al.(2005)]{su05} Su, K.~Y.~L., et al.\ 2005, \apj, 628, 487

\bibitem[Svetsov \& Shuvalov(2016)]{svetsov16} Svetsov, V.~V., \& Shuvalov, V.~V.\ 2016, \gca, 173, 50 

\bibitem[Takasawa et al.(2011)]{takasawa11} Takasawa, S., Nakamura, A.~M., Kadono, T., et al.\ 2011, \apjl, 733, L39 

\bibitem[Thebault \& Kral(2018)]{thebault18} Thebault, P., \& Kral, Q.\ 2018, \aap, 609, A98 

\bibitem[Verhoeff et al.(2012)]{verhoeff12} Verhoeff, A.~P., Waters, L.~B.~F.~M., van den Ancker, M.~E., et al.\ 2012, \aap, 538, A101 

\bibitem[Wagner et al.(2018)]{wagner18} Wagner, K., Follete, K.~B., Close, L.~M., et al.\ 2018, \apjl, 863, L8 

\bibitem[Whitney et al.(2013)]{whitney13} Whitney, B.~A., Robitaille, T.~P., Bjorkman, J.~E., et al.\ 2013, \apjs, 207, 30 

\bibitem[Winn(2018)]{winn18} Winn, J.~N.\ 2018, arXiv:1801.08543 

\bibitem[Wright et al.(2010)]{wright10} Wright, E.~L., Eisenhardt, P.~R.~M., Mainzer, A.~K., et al.\ 2010, \aj, 140, 1868-1881 

\bibitem[Wyatt et al.(2007)]{wyatt07} Wyatt, M.~C., Smith, R., Greaves, J.~S., et al.\ 2007, \apj, 658, 569 

\bibitem[Wyatt et al.(2011)]{wyatt11} Wyatt, M.~C., Clarke, C.~J., \& Booth, M.\ 2011, Celestial Mechanics and Dynamical Astronomy, 111, 1 

\bibitem[Wyatt \& Jackson(2016)]{wyatt16} Wyatt, M.~C., \& Jackson, A.~P.\ 2016, \ssr, 205, 231  

\bibitem[Zanni \& Ferreira(2013)]{zanni13} Zanni, C., \& Ferreira, J.\ 2013, \aap, 550, A99 

\bibitem[Zubko et al.(1996)]{zubko96} Zubko, V.~G., Mennella, V., Colangeli, L., \& Bussoletti, E.\ 1996, \mnras, 282, 1321 

\end{thebibliography}
\end{document}